\documentclass[11pt,draftclsnofoot,onecolumn]{IEEEtran}

\usepackage{amssymb} \usepackage{cite} \usepackage{graphicx}
\usepackage{psfrag} \usepackage{subfigure} \usepackage{url}
\usepackage{amsmath}
\usepackage{color}

\newtheorem{theorem}{Theorem} 

\newtheorem{lemma}{Lemma}
\newtheorem{claim}{Claim}
\newcommand{\remove}[1]{} 
\newcommand{\EXP}[1]{\mathsf{E}\!\left[#1\right] }
\newcommand{\Cov}[1]{\mathsf{Cov}\!\left(#1\right) }
\newcommand{\bm}[1]{\mbox{\boldmath{$#1$}}}

\newcommand{\M}[0]{\mathcal{M}}

\newcommand{\todo}[1]{\vspace{5mm}\par\noindent
{\textsc{ToDo}}
\framebox{\begin{minipage}[c]{0.85\linewidth}\tt #1
\end{minipage}}\vspace{5mm}\par}
\def\an#1{{{#1}}}

\begin{document}
  %%%%%%%%%%%%%%%%%%%%%%%%%%%%%%%%TITLE%%%%%%%%%%%%%%%%%%%%%%%%%%%%%%%%%%%%%%
  \title{Distributed and Recursive Parameter Estimation in
  Parametrized Linear State-Space Models} \author{S.~Sundhar~Ram,
  V.~V.~Veeravalli, and A.~Nedi\'{c} \thanks{The first two authors are
  with the Dept. of Electrical and Computer Engg., University of
  Illinois at Urbana-Champaign. The third author is with the Dept. of
  Industrial and Enterprise Systems Engg., University of Illinois at
  Urbana-Champaign. They can be contacted at
  \texttt{\{ssriniv5,vvv,angelia\}@uiuc.edu}. This work was partially
  supported by a research grant from Vodafone and partially by a
  research grant from Motorola.}} \maketitle
  \begin{abstract}
    We consider a network of sensors deployed to sense a
    spatio-temporal field and estimate a parameter of interest.  We
    are interested in the case where the temporal process sensed by
    each sensor can be modeled as a state-space process that is
    perturbed by random noise and parametrized by an unknown
    parameter.  \an{To estimate the unknown parameter from the
    measurements that the sensors sequentially collect, we propose a
    distributed and recursive estimation algorithm, which we refer to
    as the incremental recursive prediction error algorithm. This
    algorithm has the distributed property of incremental gradient
    algorithms and the on-line property of recursive prediction error
    algorithms. We study the convergence behavior of the algorithm and
    provide sufficient conditions for its convergence. Our convergence
    result is rather general and contains as special cases the known
    convergence results for the incremental versions of the least-mean
    square algorithm. Finally, we use the algorithm developed in this
    paper to identify the source of a gas-leak (diffusing source) in a
    closed warehouse and also report some numerical results.}
  \end{abstract}

  \section{Introduction}
  \label{sec:intro}

  \remove{With the development of the micro-fabrication technology it
  is now possible to build small low-cost sensors with on-board
  sensing, processing and wireless communication units.}  A sensor
  network consists of sensors that are spatially deployed to make
  observations about a process or field of interest. If the process
  has a temporal variation, the sensors also obtain observations
  sequentially in time. An important problem in such networks is to
  use the spatially and temporally diverse measurements collected by
  the sensors locally to estimate something of interest about the
  process. This estimation activity could either be the network's main
  objective, or could be an intermediate step such as in control
  applications where the sensors are also coupled with actuators.

  In this paper, we consider a parameter estimation problem when each
  individual sensor observation process can be modeled as a linear
  state-space process that is parametrized by an unknown parameter of
  interest, and also perturbed by process and observation
  noise. State-space models arise directly, or as linear
  approximations to non-linear models, in many applications. As an
  example, we will later discuss the problem of localizing a gas-leak
  in a warehouse.

  \an{We propose a {\it distributed and recursive} estimation
  procedure, which is suitable for in-network processing.  Each sensor
  locally processes its own data and shares only a summary in each
  time slot. The sensors form a cycle and update incrementally,
  whereby each sensor updates the estimate using its local information
  and the received estimate from its upstream neighbor, and passes the
  updated estimate to its downstream neighbor.  In this way, there is
  a reduction in communication within the network at the cost of
  increased local sensor processing. This can significantly reduce the
  total network energy used, especially, when the sensors communicate
  over a wireless medium. Furthermore, the sensor updates are
  generated recursively from every new measurement using only a
  summary statistic of the past measurements. This has two
  benefits. Firstly, the network has a (possibly coarse) estimate at
  all times, which is important in applications that require the
  network to react immediately to a stimulus and make decisions
  on-line. For example, in a network that is deployed to monitor gas
  leaks the network should raise an alert depending on the level of
  the leak intensity. An additional benefit is that each sensor can
  purge its old measurements periodically since only a summary of
  constant size is used to update the estimates. This can
  significantly reduce the memory requirements at the sensors. }

  \an{Our approach is in contrast with traditional estimation methods
  such as the maximum likelihood and least-squares which are
  centralized, i.e., the measurements collected by the spatially
  distributed sensors are routed through the network to a single
  location (fusion center) where estimates are computed. In this case,
  the network energy is mainly consumed in routing the measurements to
  the fusion center, which can be inefficient in terms of energy
  consumption.}  The problem of centralized recursive estimation in
  linear state-space models is \an{an} old problem in system
  identification. We refer the interested reader to \cite{Ljung83} for
  a survey of these methods for linear state-space models. The problem
  has also generated considerable interest in the neural networks
  community where the EM algorithm is used as a tool to learn the
  parameters \cite{Roweis99}. A related algorithm is the
  \emph{parallel recursive prediction error} algorithm proposed in
  \cite{Chen00} that updates the components of the parameter vector in
  parallel.
   
  The literature on distributed estimation is somewhat limited. A
  distributed maximum-likelihood algorithm is discussed in
  \cite{Blatt04} and a distributed expectation-maximization algorithm
  is discussed in \cite{Nowak03}.  In \cite{Rabbat04}, the incremental
  (sub)gradient algorithms of \cite{Nedic01} are used to obtain
  distributed least-square estimators.  Distributed linear
  least-squares are discussed in \cite{Xiao05} without an explicit
  point-to-point message routing.  All \an{of} these algorithms are
  distributed but not recursive.

  \an{This paper extends our earlier work \cite{Sundhar07b}}, where we
  considered the problem of recursive and distributed estimation for
  stationary models. To the best of our knowledge, there is only
  \an{one other related} study \cite{Lopes07} that deals with both
  distributed and recursive estimation. \an{There}, incremental
  versions of the least-mean square algorithm and the recursive
  least-squares are developed to solve the linear least-squares
  problem. In both \an{studies \cite{Sundhar07b} and \cite{Lopes07}},
  the models are not auto-regressive.
  
  \an{Our contribution in this paper is the development and
  convergence analysis of a general distributed recursive algorithm
  for parameter estimation in parametrized state-space models. Our
  results are more general than those of \cite{Lopes07}, which follow
  as a special case.}

  \an{The rest of the paper is organized as follows. We formulate the
    problem in Section~\ref{sec:PS}, and then introduce our notation
    in Section~\ref{sec:Notations}. We give an overview of the
    algorithm in Section~\ref{sec:AO}. We then discuss the standard
    recursive prediction error algorithm algorithm \cite{Ljung83} and
    the incremental gradient algorithm of \cite{Nedic01} in
    Section~\ref{sec:Prelim}. These algorithms are at the heart of our
    distributed algorithm presented in Section~\ref{sec:IRPE}, where
    we also state our main convergence result.  We prove the
    convergence of the algorithm in Appendix~\ref{ssec:proof}.  We
    discuss some simple extensions in Section~\ref{sec:ex}.  We report
    some experimental results obtained by our method as employed to
    localize the sourse in a gas leak problem in
    Section~\ref{sec:simul}.  We conclude in Section
    \ref{sec:conclusions}.}

  \section{Problem Formulation}
  \label{sec:PS}
  \an{We consider a network of $m$ sensors, indexed $1,\ldots,m,$
  deployed to sense a spatio-temporal diverse field to determine the
  value of some quantity of interest, denoted by $x,$ with $x \in
  \Re^d.$ We denote the true value of the parameter by $x^*$.}

  \an{We assume} that time is slotted and each sensor sequentially
  senses the field once \an{in every time slot. We model the
  measurement sequence of sensor $i$} as a random process
  $\{R_i(k;x)\}$ with the following dynamics
  \begin{align}
    \Theta_{i}(k+1;x) &= D_i(x) \Theta_{i}(k;x) + W_{i}(k;x), \cr
    R_{i}(k+1;x) &= H_i \Theta_i(k+1;x) + V_i(k+1).
    \label{eqn:statespace}
  \end{align}
  Here, $\{W_i(k;x)\}$ is the process noise, \an{$\{V_i(k)\}$ is the
  measurement noise, $H_i$ is the observation matrix and $V_i(k+1)$
  is the measurement noise of sensor $i$.
  The process $\{\Theta_{i}(k;x)\}$ can be
  interpreted as the temporal process obtained by sampling a
  spatio-temporal diverse field at the location of sensor $i.$}
  At this point, we do not assume any
  knowledge on the joint statistics of
  $\Theta_{i}(k+1;x)$ and $\Theta_{j}(k+1;x).$\footnote{Even if some 
  information is available we ignore
  it. This aspect is discussed in detail in
  Section~\ref{sec:discuss}.}

  \an{We denote by $r_{i}(k)$ the actual measurement collected by
  sensor $i$ at time slot $k,$ i.e.,} $r_{i}(k)$ is a realization of
  $R_i(k;x^*).$ The processes $\{W_{i}(k;x)\}$ and $\{V_i(k)\}$ are
  zero-mean i.i.d.\ random sequences. The quantities $D_i(x),$ $H_i,$
  $\Cov{W_i(k;x)}$ and $\Cov{V_i(k)}$ are available \an{only} at
  sensor $i.$ \an{Moreover,} at all sensors a set $X$ is available
  \an{that} satisfies the following properties
  \begin{enumerate}
  \item \an{The set} $X$ is closed and convex;
    \item \an{The true parameter $x^*$ is contained in the set $X$};
    \item \an{The system in (\ref{eqn:statespace}) is
  stable, observable and controllable for all $x \in X$}.
  \end{enumerate}
  Note that $X$ may even be the entire $\Re^d.$ The problem is to
  estimate the parameter $x$ from the collection of sensor
  measurements $\{ r_{i}(k)\}$ with an algorithm that is:
   \begin{enumerate}
  \item \emph{Distributed:} Sensor $i$ does not share its raw
    measurements $\{r_{i}(k) \}$ with \an{any other sensor}.
  \item \emph{Recursive:} At all times, sensor $i$ stores only a
    summary statistic of \an{a} constant size, i.e., the size of the
    statistic does not increase \an{with the number of measurements
    collected by the sensor}.
  \end{enumerate}

   \section{Notation}
   \label{sec:Notations}
  All the random variables are defined on the same probability space
  $\mathcal{T} = (\Omega, \mathcal{F}, \mathcal{P}).$ If $\omega \in
  \Omega,$ is the outcome of an experiment, then for a random process
  $\{Y(k;x)\}$ that is parametrized by $x,$ we define $y(k) =:
  Y_{\omega}(k;x^*)$, i.e., $y(k)$ is the value of the random variable
  $Y(k;x^*)$ corresponding to the outcome $\omega.$ According to this
  notation, $r_i(k)$ and $\theta_i(k)$ are the realizations of
  $R_i(k;x^*)$ and $\Theta_i(k;x^*)$ that correspond to the same
  outcome $\omega.$

  \an{We let $\mathcal{I}$ denote the set of sensors, i.e.,
  $\mathcal{I} := \{1,\ldots,m\}.$ } Further, we \an{assume that
  $\Theta_{i}(k;x)$ and $R_{i}(k;x)$ are vectors of dimensions $q$ and
  $p$, respectively. They are the same for all sensors $i \in
  \mathcal{I}.$\footnote{We make this assumption only for the sake of
  clarity. Our analysis applies to the general case where the
  dimensions $q$ and $p$ can be sensor dependent.}}  We write
  $R_{i}^{k}(x)$ to denote the collection of random variables
  $\{R_i(1;x),\ldots,R_i(k;x)\},$ \an{which should be viewed as a
  collection of random variables parametrized by $x$ and {\it not as a
  function of} $x.$ Furthermore, in line with our notation, $r_i^k$
  denotes the realization of $R_i^k(x^*),$ i.e., $r_i^k$ denotes the
  collection $\{r_i(1),\ldots,r_i(k)\}.$}

  \section{Algorithm Overview}
  \label{sec:AO}
  A standard estimation procedure defines the estimate as the minimum
  of a suitably defined cost that is a function of the observations
  and the unknown parameter. For example, the maximum likelihood
  estimator minimizes the negative of the log-likelihood function.
  The form of the cost function determines whether there is a
  distributed and recursive minimization procedure. Further, the cost
  function also determines other properties of the estimator such as
  unbiasedness, consistency, minimum variance etc. (see
  \cite{Poor94}).

  Except in some very special estimation problems, it is impossible to
  find a cost function that supports even a centralized recursive
  procedure and also generates a `good' estimate.  In this paper, we
  develop a distributed and recursive estimator that is only
  \emph{consistent}, i.e., the estimate converges to the correct value
  $x^*$ as the number of available measurements becomes
  infinite\footnote{This statement is technically imprecise and will
  be clarified later.}. The estimates are biased but this is the price
  that is to be paid to obtain a distributed and recursive
  procedure. Thus, there are two aspects to the problem. The first is
  to choose a suitable cost function, and the second is to develop a
  distributed and recursive minimization procedure. We will first
  discuss the cost function that is used, and then give an overview of
  the minimization algorithm.

  \subsection{Cost function}
  \label{ssec:Kalman}
  Suppose that each sensor has made $N$ measurements and we want to
  estimate the parameter $x$ from these measurements. As mentioned,
  the cost is a function of both the available measurements and the
  unknown parameter.  Therefore, we denote the cost function as
  $f_N(x;r^{N}).$

  For $x\in X,$ we assumed that the system in (\ref{eqn:statespace})
  is stable, observable and controllable. The Kalman gain for the
  system therefore converges to a finite time-invariant value
  \cite{Kumar86}.  Let $G_{i}(x)$ be the Kalman gain for the
  state-space system in (\ref{eqn:statespace}), which is determined
  from $D_i(x), H_i,$ $\Cov{W_i(k;x)},$ and $\Cov{V_i(k)}$ as the
  solution to the Riccati equation \cite{Ljung83}. Using $G_i(x)$
  define
  \begin{align}
    \phi_{i,k+1}(x;r_{i}^{k}) =& \left( D_i(x) - G_i(x) \right)
    \phi_{i,k}(x;r_i^{k-1}) \cr
    &+ G_{i}(x) r_{i}(k),\cr g_{i,k+1}(x;r_i^{k}) =& H_i
    \phi_{i,k+1}(x;r_{i}^{k}),\label{eqn:Kalman_s}
  \end{align}
  with $\phi_{i,1}(x;r_{i}^0) = \mu_{i}(x).$ \an{Observe that
    $g_{i,k+1}(x;r_{i}^{k})$ is linear in $r_{i}^{k}$ for each $x.$
    Furthermore, for} any $x \in \Re^d,$ $g_{i,k+1}(x;r_{i}^{k})$
  viewed as a function of $r_{i}^{k}$ is an one-step prediction
  filter (henceforth, \an{referred to} as a predictor) for the
  random process $\{R_i(k+1;x^*)\}.$ Thus,
  $\{\hat{g}_{i,k+1}(x;r_{i}^{k})\}_{x\in \Re^d}$ is a predictor
  family parametrized by $x.$

  We will choose our cost function to be
  \begin{align}
    f_{N}(x;r^{N}) = \frac{1}{N} \sum_{k=1}^{N} \sum_{i=1}^{m}
    \left\|r_{i}(k) - g_{i,k}(x;r_i^{k-1}) \right\|^2, \label{eqn:cf}
  \end{align} 
  and our estimator to be
  \begin{align*}
    \hat{x}_{N} = \arg \min f_{N}(x;r^{N}).
  \end{align*}
  We next provide an intuitive explanation as to why this choice of
  cost function should generate consistent estimates. First, note that
  the vector $g_{i,k+1}(x^*;r_{i}^{k})$ is the steady-state Kalman
  predictor for $R_{i}(k+1;x^*)$ since $\{r_{i}(k)\}$ is a sample path
  of the random process $\{R_i(k;x^*)\}.$ Among other properties, the
  steady state Kalman filter is asymptotically optimal in a mean
  square sense in the class of linear time-invariant predictors. Thus,
  $x^*$ will minimize
  \begin{align*}
      f(x) &= \lim_{N \to \infty} \frac{1}{N} \sum_{k=1}^{N}
      \sum_{i=1}^{m} \mathsf{E}\left[ \left\| R_{i}(k;x^*)
      -g_{i,k}(x;R_i^{k-1}(x^*)) \right\|^2\right] \\&=
      \EXP{f_{N}(x;R^{N}(x^*))}.
  \end{align*}
  Here, the expectations are taken with respect to the random process
  $\{R_i^{k}(x^*)\}.$ Therefore, one can expect that $\hat{x}_{N},$
  which is the minimum of $f_{N}(x;r^{N}),$ might in the limit be
  equal to $x^*,$ since $x^*$ is the minimum of $f(x),$ the limit of
  $\EXP{f_{N}\left(x;R^{N}(x^*)\right)}.$

  \subsection{Overview of the minimization procedure}
\begin{figure}[tb]
    \center{\includegraphics[scale = 0.45]{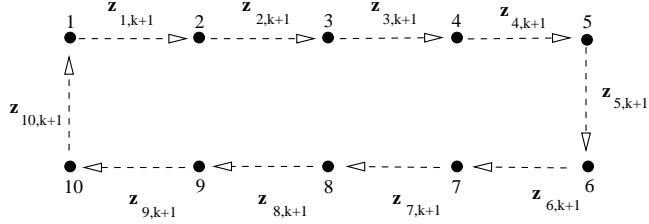}}
    \caption{A network of $10$ sensors with incremental
      processing. The estimate is cycled through the network. The
      quantity $z_{i,k+1}$ is the intermediate value \an{after the sensor $i$
	update at time $k+1$}.}
    \label{Fig:cycle}
  \end{figure}

  Observe that $f_N(x,r^N)$ can be written as
   \[ 
    f_{N}\left(x;r^{N}\right) = \sum_{i=1}^{m}
   f_{i,N}\left(x;r_i^{N}\right),
   \] 
   where 
   \[
   f_{i,N}\left(x;r_i^{N}\right) =: \frac{1}{N} \sum_{k=1}^{N}
   \left\|r_{i}(k) - g_{i,k}(x;r_i^{k-1}) \right\|^2.
   \]
   Suppose we are interested in a non-recursive but distributed
   solution the problem. Then, the incremental gradient algorithm
   \cite{Nedic01} can be used to minimize $f_N(x;r^N).$ In each time
   slot, the algorithm cycles the estimate through the sensor network.
   Sensor $i$ receives the estimate $z_{i-1,k+1}$ from sensor $i-1$ at
   time slot $k+1$, and generates a new estimate $z_{i,k+1}$ using
   $\nabla f_i(z_{i-1,k+1})$.  The new estimate is then passed to
   sensor $i+1$ for $i<m,$ and to sensor $1$ for $i=m$ and, thus, the
   estimate is cycled through the network for each sensor to
   update. An illustration is shown in Fig.\,\ref{Fig:cycle}.

   Now consider the complementary problem, i.e., a centralized but
   recursive solution. By recursive we mean, the algorithm should be
   able to obtain $\hat{x}_{N+1}$ directly from $\hat{x}_{N},$ the new
   measurements $r_{1,N+1}, \ldots,r_{m,N+1},$ and some summary
   statistic of the past observations. This is generally not
   possible. Nevertheless, it is possible to use the recursive
   prediction error algorithm of \cite{Ljung83} to obtain recursive
   approximations to $\{\hat{x}_{k}\}$ such that the approximate
   sequence converges to the same limit as $\{\hat{x}_{k}\}.$ Thus,
   while we do not minimize the chosen cost function $f_N$ at each
   step, the new sequence $\{x_{k}\}$ is still consistent.
   
   The algorithm that we propose is a combination of the incremental
   gradient algorithm and the recursive prediction error algorithm. We
   therefore refer to the algorithm developed in this paper as the
   \emph{incremental recursive prediction error} algorithm.

  \section{Preliminaries}
  \label{sec:Prelim}
  We first evaluate some quantities that will be useful to us in the
  analysis.  \an{To make the paper self-contained, we then discuss the
  incremental gradient method \cite{Nedic01} and the recursive
  prediction error algorithm \cite{Ljung83} in this section.}
  
  \subsection{Some more notation}
  For later reference, we obtain the form of the gradient of the
  predictor $g_{i,k+1} \left(x;r_i^k \right).$ Define $F_i(x) =
  D_{i}(x) - G_{i} (x) H_{i}$ and for convenience rewrite
  (\ref{eqn:Kalman_s}) as follows
  \begin{align}
    \phi_{i,k+1}(x;r_{i}^{k}) =& F_i(x) \phi_{i,k}(x;r_i^{k-1}) +
    G_i(x) r_{i}(k), \nonumber \\ g_{i,k+1}(x;r_i^{k}) =& H_i
    \phi_{i,k+1}(x;r_{i}^{k}).
    \label{eqn:detmodel}
  \end{align}
  Let $x^{(\ell)}$ denote the $\ell$-th component of $x,$ 
  \an{and define}
   \begin{eqnarray*}
     \zeta^{(\ell)}_{i,k}(x;r_{i}^{k-1}) &= \frac{ \partial
       \phi_{i,k}(x;r_i^{k-1})}{\partial x^{(\ell)}}, \ \
     \nabla^{(\ell)} F_i(x) &= \frac{
       \partial F_i(x) }{\partial x^{(\ell)}}, \\
     \eta^{(\ell)}_{i,k}(x;r_{i}^{k-1}) &= \frac{ \partial
       g_{i,k}(x;r_i^{k-1})}{\partial x^{(\ell)}},  \ \ 
	\nabla^{(\ell)} G_i(x) &= \frac{
	  \partial G_i(x) }{\partial x^{(\ell)}}.
   \end{eqnarray*}
   Thus the gradient $\nabla g_{i,k}(x;r_{i}^{k-1})$ is the $p \times
   d$ matrix,
   \begin{align}
     \nabla g_{i,k}(x;r_{i}^{k-1}) = \left[
       \eta^{(1)}_{i,k}(x;r_{i}^{k-1}) \ \ \ldots \ \
   \eta^{(d)}_{i,k}(x;r_{i}^{k-1}) \right]. \label{eqn:gradform}
   \end{align}
   By differentiating in (\ref{eqn:detmodel}), we can immediately see
   that 
   \begin{align}
     \left[ \begin{array}{c} \phi_{i,k+1}(x;r_{i}^{k}) \\
	 \zeta^{(\ell)}_{i,k+1}(x;r_{i}^{k})
       \end{array}  \right]
     =&  \left[ \begin{array}{cc}
	 F_i(x) & 0 \\
  \nabla^{(\ell)} F_i(x) & F_i(x) 
       \end{array}
       \right]
     \left[ \begin{array}{c} \phi_{i,k}(x;r_{i}^{k-1}) \\
	 \zeta^{(\ell)}_{i,k}(x;r_{i}^{k-1})
       \end{array} 
       \right] \nonumber 
     + 
     \left[ \begin{array}{c}
	 G_i(x) \\
	 \nabla^{(\ell)} G_i(x)
	  \end{array} 
       \right] 
	 r_{i}(k), 
     \nonumber
     \\ 
     \left[
       \begin{array}{c}
	 g_{i,k+1}(x;r_i^{k}) \\ \eta^{(\ell)}_{i,k+1}\left(x;r_i^k \right)
       \end{array}
       \right] =& \left[ \begin{array}{cc}
	 H_i & 0 \\
	 0   & H_i
       \end{array} 
       \right]      \left[ \begin{array}{c}
	 \phi_{i,k+1}(x;r_{i}^{k}) \\ \zeta^{(\ell)}_{i,k+1}(x;r_{i}^{k})
       \end{array}  \right].
     \label{eqn:derivatives}
   \end{align}

  \subsection{Incremental gradient descent algorithm}
  \an{For differentiable optimization problem of the form} 
  \begin{eqnarray*}
    \min_{x \in X} \sum_{i=1}^{m} f_{i}(x),
  \end{eqnarray*}
  \an{the standard gradient descent method, with projections, generates 
    iterates according to the following rule:}
  \begin{eqnarray*}
    x_{k+1} &=& \mathcal{P}_{X} \left[ x_{k} - \alpha_{k+1}
      \sum_{i=1}^{m} \nabla f_i(x_{k}) \right]. 
  \end{eqnarray*}
  Here, \an{the scalar $\alpha_{k+1}>0$} is the step-size and
  $\mathcal{P}_{X}$ denotes \an{the} projection onto the set $X.$
  \an{This method is centralized in the sense that it requires the
  gradient information of each $f_i(x)$ at the current iterate $x_k$
  in order to generate the new iterate $x_{k+1}$. In our setting,
  however, the gradient information $\nabla f_i(x)$ is distributed
  since $f_i(x)$ is known only locally at sensor $i.$ Thus, the
  standard gradient descent method is not adequate.}
   
  \an{To deal with the distributed nature of the sensor network
    information, we consider the incremental gradient method to}
    minimize $f(x),$ without the sensors explicitly sharing the
    functions $f_i(x)$ (see, \cite{Nedic01}, \cite{Bertsekas00} and
    the references therein). In this algorithm, the iterates are
    generated according to
  \begin{eqnarray}
    x_{k} &=& z_{m,k} = z_{0,k+1}, \cr 
    z_{i,k+1} &=&
    \mathcal{P}_{X} \left[z_{i-1,k} - \alpha_{k+1} \nabla f_{i}(
      z_{i-1,k})\right]. \label{eqn:incgrad}
  \end{eqnarray}
  \an{The key difference between the standard gradient and incremental
    gradient method is that the standard gradient method generates
    iterates by using the gradient information of all functions
    $f_{i}(x)$ at the same (current) estimate $x_k$, while the
    incremental method generates iterates through a cycle of
    intermittent adjustments $z_{i-1,k+1}$ using only one function at
    a time, i.e., the gradient $\nabla f_i(z_{i-1,k+1})$, so that all
    functions $f_i$ are processed within a cycle (see
    Fig.\,\ref{Fig:cycle} for an illustration).}  The convergence of the
    incremental gradient method has been studied \an{in
    \cite{Bertsekas00}, \cite{Nedic01},} \cite{Solodov98} under
    different assumptions on the functions $f_i(x)$ and the step-size
    rules.

   \subsection{Recursive prediction error algorithm}
   \an{Here, we discuss the standard recursive prediction error algorithm 
   (RPE) for a parameter estimation problem (see \cite{Ljung83}).
   To avoid confusion with the notation in the rest of the paper, 
   we suppress the subscript $i$ and consider the problem of
   estimating $x$ from observations of a random process $\{R(k;x)\}$ 
   with the following dynamics:}
   \begin{align}
     \Theta(k+1;x) &= D(x) \Theta(k;x) + W(k;x), \nonumber
     \\ R(k+1;x) &= H \Theta(k+1;x) + V(k).
     \label{eqn:ss1}
   \end{align}
   \an{The RPE algorithm generates estimates of $x$ by applying
   suitable approximations to the iterates generated by the gradient
   descent method as employed to solve an appropriate optimization
   problem.}  \an{In particular,} on the set $X,$ the true
   \an{parameter value $x^*$ minimizes the following function:}
   \begin{equation}
     f(x) = \lim_{N \to \infty} \frac{1}{N} \sum_{k=1}^{N}
     \EXP{\left\| R(k;x^*) -
       g_{k}(x;r^{k-1}(x^*))
       \right\|^2}. \label{eqn:costfnex}
   \end{equation}
   \an{When} the standard gradient descent method is used to minimize $f,$
   the iterates $\hat{x}_{k+1}$ \an{are given by}
   \begin{align}
     \hat{x}_{k+1} = \mathcal{P}_{X} \left[\hat{x}_k - \alpha_{k+1}
       \nabla f(\hat{x}_k)\right]. \label{eqn:graddesc}
   \end{align}
   The RPE algorithm obtains a sequence $\{x_k\}$ by using two
   approximations to the sequence $\{\hat{x}_k\}.$ 

   \an{The gradient of $f(x)$ is not available in (\ref{eqn:graddesc}),
   but instead the sequence $\{r(k)\}$ is available. The first
   approximation is a least mean-square (LMS) type approximation 
   replacing the actual
   gradient $\nabla f(\hat{x}_k)$ with an empirical gradient. 
   Let us denote the iterate sequence corresponding to this approximation
   by $\{\bar{x}_k \}$, for which  we have}
   \[
   \bar{x}_{k+1} =\mathcal{P}_{X} \left[\bar{x}_k - \alpha_{k+1} \nabla
   \hat{f}_{k+1}(\bar{x}_k;r^{k+1})\right],
   \]
   where
     \begin{align*}
       \nabla \hat{f}_{k+1}(\bar{x}_k;r^{k+1}) &= - 2 \ \left(\nabla
       g_{k+1}(\bar{x}_k;r^k)\right)^T (r(k+1) -
       g_{k+1}(\bar{x}_k;r^k) ).
     \end{align*}
   The gradient $\nabla g_{k+1}(x;r^k),$ can be obtained from
   (\ref{eqn:gradform}) and the extended representation of
   $\eta^{(\ell)}_{k+1}(x;r^k)$ in (\ref{eqn:derivatives}).  The problem
   is that even with this approximation the sequence $\{ \bar{x}_k\}$
   cannot be obtained recursively. Observe that to exactly evaluate
   $g_{k+1}(\bar{x}_k;r^{k})$ and $\nabla g_{k+1}(\bar{x}_k;r^{k})$
   one would need the entire vector $r^{k}.$ \an{To accommodate the
   recursive computations,} we use another approximation
   \begin{align}
     \left[ \begin{array}{c} 
	\phi_{k+1} (\bar{x}_k;r^{k}) \\ 
	\zeta^{(\ell)}_{k+1} (\bar{x}_k;r^{k})
     \end{array}  \right] 
   &\simeq \left[ \begin{array}{cc}
	 F(\bar{x}_k)  & 0 \\
	 \nabla^{(\ell)} F(\bar{x}_k)  & F(\bar{x}_k) 
       \end{array}
       \right] \left[ \begin{array}{c} \phi_{k}
     (\bar{x}_{k-1};r^{k-1}) \\
     \zeta^{(\ell)}_{k}(\bar{x}_{k-1};r^{k-1})
     \end{array} 
     \right]\nonumber + \left[ \begin{array}{c} G(\bar{x}_k) \\
	 \nabla^{(\ell)} G(\bar{x}_k)
	  \end{array} 
       \right]  r(k), \nonumber \\ \left[
     \begin{array}{c}
       g_{k+1}(\bar{x}_k;r^{k}) \\
       \eta^{(\ell)}_{k+1}(\bar{x}_k;r^{k})
     \end{array}
     \right] &= \left[ \begin{array}{cc} H & 0\\
	 0 & H \end{array}
	 \right] \left[ \begin{array}{c}
     \phi_{k+1} (\bar{x}_{k};r^{k}) \\ \zeta^{(\ell)}_{k+1}
     (\bar{x}_k;r^{k})
     \end{array}  \right]. \label{eqn:approx}
   \end{align}
   Changing notation to reflect the approximations and re-ordering
   the equations, the \an{resulting} RPE algorithm can be written as
   \an{follows, for $\ell=1,\ldots,d,$}
   \begin{eqnarray}
    \left[
       \begin{array}{c}
		 h_{k+1}\\ 
		 \xi^{(\ell)}_{k+1}
       \end{array}
       \right] &=& \left[ \begin{array}{cc} H & 0\\
	 0 & H \end{array}
	 \right] \left[ \begin{array}{c} 
		 \psi_{k+1} \\ 
		 \chi^{(\ell)}_{k+1}
       \end{array}  \right],  \nonumber \\
     \epsilon_{k+1} &=& r(k+1) - h_{k+1}, \nonumber \\
     \underline{x}^{(\ell)}_{k+1} &=& x^{(\ell)}_k - \alpha_{k+1}\
     \left(\xi^{(\ell)}_{k+1}\right)^T \epsilon_{k+1}, \nonumber \\
     \underline{x}_{k+1} &=& \left[ \underline{x}^{(1)}_{k+1} \ \
     \ldots \ \ \underline{x}^{(d)}_{k+1} \right]^T, \nonumber
     \\ 
     x_{k+1} &=& \mathcal{P}_{X} \left[ \underline{x}_{k+1} \right], 
     \nonumber\\
     \left[
     \begin{array}{c} \psi_{k+2} \\ \chi^{(\ell)}_{k+2}
       \end{array}  \right]
     &=&  \hspace{-0.1 in} \left[ \begin{array}{cc}
	 F(x_{k+1})  & 0 \\
	 \nabla^{(\ell)} F(x_{k+1})  & F(x_{k+1}) 
       \end{array}
       \right]
      \left[ \begin{array}{c} \psi_{k+1} \\ 
     \chi^{(\ell)}_{k+1}
       \end{array}  \right] +    
      \left[ \begin{array}{c}
	 G(x_{k+1}) \\
	 \nabla^{(\ell)} G(x_{k+1})
	  \end{array} 
       \right]
      r(k+1). 
     \label{eqn:RPE} 
   \end{eqnarray}
   %Here $l = 1,\ldots,d.$ 
   The algorithm is initialized with values for
   $\psi_1, \chi^{(\ell)}_{1}$ and $x_{0}.$ Observe that to update $x_k$
   the algorithm requires \an{only} $r(k+1),$
   $\chi^{(1)}_{k+1},\ldots,\chi^{(d)}_{k+1}$ and $\psi_{k+1},$
   \an{and therefore, it} is recursive.

   In summary, the iterates of the RPE algorithm are
   obtained from the standard gradient descent iterates
   with the following two approximations:
   \begin{enumerate}
   \item \an{An} LMS-like approximation for the gradient, and 
   \item An approximation to make the LMS approximations recursive.
   \end{enumerate}
   \an{The following theorem provides some sufficient 
   conditions guaranteeing that the iterates generated by
   the RPE algorithm asymptotically converge to a minimum of $f(x).$
   The theorem is based on the results from \cite{Ljung83}.} 
   \emph{
     \begin{theorem}
       \label{thm:rpe}
       Let the following conditions hold.
       \begin{enumerate}
       \item The set $X$ is a closed and convex set containing $x^*.$
	 \an{Furthermore, the system in (\ref{eqn:ss1}) is
	 stable, observable and controllable for all $x \in X$}.
       \item The matrices $F(x)$ and $G(x)$ 
         are twice differentiable
	 for all $x \in X.$
       \item The fourth moments of $V(k)$ are bounded. The second
	 moments of $W(k;x^*)$ are bounded.
       \end{enumerate}
       \an{Moreover}, let the step-size $\alpha_k$ be 
       such that $k \alpha_k \to \mu$ \an{for some positive scalar $\mu$}. 
       Then, the iterates $x_k$ generated by the RPE
       in (\ref{eqn:RPE}) converge to a local minimum of $f(x)$ in
       (\ref{eqn:costfnex}) over the set $X$, with probability $1.$
     \end{theorem}
   } Theorem~\ref{thm:rpe} follows from Theorem~4.3 on page~182 and
   the discussions in pages~172~and~184 of \cite{Ljung83}. The
   conditions for convergence of the algorithm are extremely
   weak. Note that the algorithm guarantees convergence only to a
   local minima and not necessarily to the global \an{minimum} $x^*$
   of $f(x).$ Of course, when the function $f(x)$ is convex this
   implies convergence to a global minimum.

  \section{Incremental Recursive Prediction Error Algorithm}
  \label{sec:IRPE}
  As discussed in Section~\ref{ssec:Kalman}, when there are multiple
  sensors, \an{the true parameter $x^*$ minimizes}
  \begin{align}
   f(x) &= \sum_{i=1}^{m} \lim_{N \to \infty} \frac{1}{N}
  \sum_{k=1}^{N} \EXP{\left\| R_{i}(k;x^*) - g_{i,k}(x;r_i^{k-1}(x^*))
  \right\|^2} \nonumber \\ &= \sum_{i=1}^{m}
  f_i(x). \label{eqn:costfn}
  \end{align}
  We combine the incremental gradient algorithm in
  (\ref{eqn:incgrad}) with the RPE algorithm in (\ref{eqn:RPE}) to
  develop an incremental recursive prediction error (IRPE) algorithm.
  The main idea of the IRPE is to use \an{an} RPE like approximation for the
  gradient term in the incremental gradient algorithm
  (\ref{eqn:incgrad}).  \an{Formally}, the iterates are generated
  according to \an{the following relations 
  for $i\in\mathcal{I},$ and $\ell=1,\ldots,d,$} 
  \begin{eqnarray}
   x_{k} &=& z_{m,k} = z_{0,k+1}, \nonumber \\
    \left[
      \begin{array}{c}
	h_{i,k+1}\\ 
	\xi^{(\ell)}_{i,k+1}
      \end{array}
      \right] &=&  \left[ \begin{array}{cc} H_{i} & 0\\
	0 & H_{i} \end{array}
	\right] \left[ \begin{array}{c} 
	\psi_{i,k+1} \\ 
	\chi^{(\ell)}_{i,k+1}
      \end{array}  
      \right], \label{eqn:IRPE1} \\ \epsilon_{i,k+1} &=& r_{i}(k+1)-
     h_{i,k+1}, \label{eqn:IRPE2} \\ \underline{z}^{(\ell)}_{i,k+1} &=&
     z^{(\ell)}_{i-1,k+1} - \alpha_{k+1} \ \left( \xi^{(\ell)}_{i,k+1}
     \right)^T \epsilon_{i,k+1},
    \label{eqn:IRPE3} \\
    \underline{z}_{i,k+1} &=& \left[ \underline{z}^{(1)}_{i,k+1} \ \ \ 
      \ldots \ \ \
     \underline{z}^{(d)}_{i,k+1} \right]^T, \label{eqn:IRPE10} \\
    z_{i,k+1}&=& \mathcal{P}_{X} [\underline{z}_{i,k+1} ], 
    \label{eqn:IRPE11} \\
    \left[ \begin{array}{c} \psi_{i,k+2} \\ \chi^{(\ell)}_{i,k+2}
      \end{array}  \right]
    &=& \hspace{- 0.15 in}
    \left[ \begin{array}{cc}
	F_{i}(z_{i,k+1})  & 0 \\
	\nabla^{(\ell)} F_{i}(z_{i,k+1})  & F_{i}(z_{i,k+1}) 
      \end{array}
      \right]  \left[ \begin{array}{c} 
	\psi_{i,k+1} \\ 
	\chi^{(\ell)}_{i,k+1} 
      \end{array}  \right] 
    + \left[ \begin{array}{c}
	G_{i}(z_{i,k+1}) \\
	\nabla^{(\ell)} G_{i}(z_{i,k+1})
	 \end{array} 
      \right] 
    r_{i}(k+1). \label{eqn:IRPE4}
  \end{eqnarray}
  The initial values for the recursion are fixed at $x_0 = x_{s},$
  $\psi_{i,1} = \psi_{i,s}$ and $\chi^{(\ell)}_{i,1} = \chi^{(\ell)}_{i,s}.$
  To see that the algorithm has a distributed and recursive
  implementation assume sensor $i-1$ communicates $z_{i-1,k+1}$ to
  sensor $i$ in slot $k+1.$ Sensor $i$ then uses\footnote{We are
  assuming that sensor $i$ obtains its measurement before it receives
  the iterate. From an implementation perspective, each time slot can be
  divided into two parts. In the first part, the sensors make
  measurements and in the second part they process.} $r_{i}(k+1)$ to
  updates the iterate $z_{i-1,k+1}$ to generate $z_{i,k+1}.$ This is
  then passed to the next sensor in the cycle.  Observe that in
  updating $z_{i-1,k+1},$ sensor $i$ requires only
  $\chi^{(1)}_{i,k+1}, \ldots \chi^{(d)}_{i,k+1}$ and $\psi_{i,k+1},$
  \an{which} were calculated by sensor $i$ in the previous \an{time} slot. 
  Thus, the algorithm is recursive and distributed. 
  \an{Furthermore}, note that sensor
  $i$ only needs to know its own system matrices $H_i, F_i(x)$ and
  $G_i(x).$

  \subsection{Convergence result}
  The iterates generated by the IRPE \an{method} are three
  approximations away from the iterates generated by the standard
  gradient descent method. The first approximation is in going from
  the standard gradient algorithm to the incremental gradient
  algorithm, the second is in approximating the gradient of the
  function with an LMS-like empirical gradient and the third is in
  calculating the empirical gradient recursively. Therefore, it is not
  clear if the iterates will converge to $x^*.$ We next \an{state} a
  theorem that provides sufficient conditions for the convergence of
  the iterates generated by the IRPE \an{algorithm}.  \emph{
    \begin{theorem}
      \label{thm:IRPE}
      For all $i \in \mathcal{I},$ let the following conditions hold
      \begin{enumerate}
      \item The set $X$ is a closed and convex set containing $x^*.$
      \an{Furthermore, the system in (\ref{eqn:ss1}) is
      stable, observable and controllable for all $x \in X.$}
      \item The matrices $F_i(x)$ and $G_i(x)$ are twice
      differentiable for all $x \in X$.
      \item The fourth moments of $V_i(k)$ are bounded. The second
	moments of $W_i(k;x^*)$ are bounded.
      \end{enumerate}
      \an{Moreover, let the step-size $\alpha_k$ be such that $k
      \alpha_k \to \mu$ for some positive scalar $\mu$.}  Then, the
      iterates $x_{k}$
      generated by the IRPE \an{algorithm in
      (\ref{eqn:IRPE1})--(\ref{eqn:IRPE4}) converge} to a local
      minimum of $f(x)$ in (\ref{eqn:costfn}) over the set $X$,
      \an{with probability 1}.
    \end{theorem}
  }

  Note that the result implies that for each $i$ the iterates
  $z_{i,k+1}$ converge to the same local minimum.  Thus the algorithm
  is not necessarily consistent.

  \an{There is alternative way to interpret the IRPE algorithm. In
  particular,} consider a centralized scheme where the sensors
  immediately communicate their measurements to a fusion center. Now,
  at the fusion center, the RPE algorithm can be used to estimate the
  parameter $x.$ For the specific model of \an{our} interest, there is
  a hidden structure \an{in} the RPE algorithm that \an{permits an
  incremental implementation.  Thus, the IRPE algorithm can also be
  viewed as an incremental implementation of a centralized RPE
  algorithm. Since this hidden structure in the RPE algorithm is not
  easily identified, we have not used this alternative approach to
  actually present the algorithm.  However, we use this approach to
  prove the convergence result stated in Theorem \ref{thm:IRPE}. The
  proof is provided in Appendix~\ref{ssec:proof}.}

  \subsection{Communication requirements} 
  Incremental algorithm can potentially require less communication
  than centralized schemes. In a centralized scheme, in every slot
  each sensor has to communicate its measurements to a fusion center
  that is $O(1)$ meters away \an{on average}. Summed over the $m$
  sensors in the network, the total communication requirement in a
  centralized scheme is $O(m)$ bit meters per slot.  In the
  incremental scheme, each sensor needs to pass only the iterate to a
  neighbor which is $O\left(\frac{\log m}{\sqrt{m}} \right)$ meters
  away on average, as discussed in \cite{Rabbat04}.  Therefore, the
  total communication required in the incremental processing scheme is
  $O(\sqrt{m} \log m )$ bit meters per slot.

  \subsection{Centralized versus incremental: Tradeoff}
  \label{sec:tradeoff}
  In our analysis we do not use any information about the joint
  statistics of the random process $\{\Theta_i(k;x)\}$ and
  $\{\Theta_j(k;x)\}.$ When this is the case the performance of the
  IRPE is identical to the performance of the centralized RPE
  algorithm.

  Suppose some information about the join statistics is available.
  This information \emph{cannot} be used in a distributed system
  because at most one sensor's measurement is known at a single
  location at any time. Thus, the joint distribution information is
  not useful to the RPE.

  A centralized system, on the other hand, can potentially use the
  joint density information to obtain a cost function $f(x)$ that
  generates estimates with better properties. \an{As an example,
  suppose that $\Theta_i(k;x) = \Theta_j(k;x)$ for all $k\ge1$} and
  $i,j \in \mathcal{I},$ \an{which corresponds to} the case when
  \an{all sensors} sense a field with no spatial variation,
  synchronously at time $mk$.  Define $H,$ respectively\ $V(k+1),$ to
  be the block matrix obtained by stacking \an{the matrices
  $H_1,\ldots,H_m,$ respectively vectors
  $V_{1}(k+1),\ldots,V_m(k+1).$} Then, the centralized measurements
  $R(k;x)$ \an{have the following evolution:}
  \begin{align} 
    \Theta(k+1;x) &= D(x) \Theta(k;x) + W(k;x), \nonumber \\ 
    R(k+1;x)   &= H \Theta(k+1;x) + V(k+1). \label{eqn:Css}
  \end{align}
  \an{The corresponding time-invariant Kalman predictor is given by}
  \begin{align*} 
    \phi_{k+1}(x;r^k) &= \left(D(x) - G(x) H\right) \phi_k(x;r^{k-1})
    + G(x) r(k), \nonumber \\ g_{k+1}(x;r^{k}) &= H \phi_{k+1}(x;r^k).
  \end{align*}
  Notice that the \an{predictor $g_{i,k+1}(x;r^{k})$} for the
  $(k+1)$-\an{st measurement of sensor $i$} is a function of the past
  measurements made by sensor $j,$ $j \neq i.$ Using this predictor we
  can define a cost function in a manner similar to (\ref{eqn:cf}).
  As we will see in Section~\ref{sec:simul}, the nature of the cost
  function may be significantly better in terms of the number of local
  minima and the RPE applied to the system in (\ref{eqn:Css}) may have
  a better performance.

  To summarize, there is an implicit tradeoff when we use the
  IRPE. Potentially better estimates may be obtained by a centralized
  scheme when the joint statistics of the process $\Theta_i(k;x)$ and
  $\Theta_j(k;x)$ are available. This \an{is indicated by our}
  numerical results in Section~\ref{sec:simul}.

  \section{Extensions}
  \label{sec:ex}
  We next discuss some extensions to the IRPE algorithm.
  \subsection{Hybrid scheme}
  Let us consider an alternative network architecture where the
  network of $m$ sensors, divided into $m_c$ clusters of approximately
  equal size, is deployed in a unit square. Each cluster has a cluster
  head that is a neighbor to all the sensors in the cluster. We can
  develop a hybrid algorithm that is centralized intra-cluster and
  distributed inter-cluster. Each cluster head collects all the
  measurements made by the sensors in the cluster, and then the
  cluster heads use the IRPE algorithm to estimate $x$ without sharing
  their measurements.

  Note that, as each sensor is in the neighborhood of its cluster
  head, it is still required to only communicate to a neighbor. The
  cluster heads might have to communicate over larger distances. The
  total inter-cluster communication is $O(m_c)$ bits per meter and the
  total communication in a cluster is $O\left(\sqrt{\frac{m}{m_c}}
  \log\left(\frac{m}{m_c}\right)\right)$ bits over an average distance
  of $\frac{1}{m_c}.$ Therefore, the total communication is
  $O\left(\sqrt{\frac{m}{m_c}} \log\left(\frac{m}{m_c}\right)\right) +
  O(m_c)$ bits per meter. The benefit is that the cluster heads can
  use any information that is available about the joint statistics of
  the processes seen by the sensors in the cluster. 

  \subsection{Distributed and recursive regression}
  \label{sec:discuss}
  In the problem we have studied, the actual sensor measurement
  \an{sequence} $\{r_{i}(k)\}$ is a sample path of the random process
  $\{R_{i}(k;x)\}$ \an{for} $x = x^*.$ While we did \an{not assume to}
  know the value of $x^*,$ we did assume that for some $x \in X$ the
  actual system is correctly modeled by (\ref{eqn:statespace}).

  In practice, it is very difficult to obtain the correct description
  and often approximate models are used. In the context of \an{our}
  problem, this means that (\ref{eqn:statespace}) need not necessarily
  be the correct description of the actual system dynamics for any
  value of $x \in X.$ The minimum of $f(x)$ is now interpreted as the
  value of $x$ that chooses the state-space system \an{that best
  approximates the actual system among all the systems generated as
  $x$ ranges over $X.$}

  Theorem~4.3 of \cite{Ljung83} concludes that even in this case,
  under some weak regularity conditions on the actual measurement
  sequence (instead of Condition~3) the iterates generated by the RPE
  still converge to a local minimum of $f(x).$ Since the IRPE was
  proved to be equivalent to the centralized RPE, the above statement
  extends to the IRPE algorithm also.
   
  \subsection{Other extensions}
  We have not included an explicit input in modeling the system. Much
  of the analysis immediately follows where there is a deterministic
  open-loop input $\{u_i(k)\}$ that drives the system in
  (\ref{eqn:statespace}). Of course, $\{u_i(k)\}$ should be known to
  sensor $i.$ Another immediate extension is to the case when \an{the
  matrix $H_i$ and noise} $V_i(k)$ are also be parametrized by $x.$
  Finally, we remark that rate of convergence results are available
  for the RPE through a central limit theorem and these can be
  extended to the IRPE.

  \section{Application}
  \label{sec:simul}
  We next \an{consider a gas-leak} problem to illustrate the concepts
  developed in the paper.  \an{We assume that a wireless sensor
  network is deployed inside a warehouse where gas tanks are
  stored. The network objective is to localize a leak, when one
  occurs. We use a two-dimensional model for this scenario, which is
  appropriate when the gas is significantly heavier than air. In any
  case, the extension to three dimension is immediate.} We also remark
  that we have used the gas leak problem as only a representative
  example; the analysis is more generally applicable to heat and other
  diffusing sources.

  \subsubsection*{Leak model}
  \an{We asume that the leak occurs at time $t=0$ and that the network
  detects the leak immediately.} We model the leak as a point source
  at $x=(x_1,x_2).$ Each sensor sampling \an{has a duration of 1 time
  unit. The leak intensity is modeled as a piece-wise constant
  function, i.e., the leak intensity is equal to $I_{k}$ during the
  time interval $[k-1,k)$ for $k \ge 1.$} Across sampling intervals,
  the \an{leak} intensity values vary according to the following
  Markov process:
  \begin{equation}
    \label{eqn:gmp}
    I(k+1) = \rho I(k) + S(k).
  \end{equation}
  Here, $\rho$ is a known \an{scalar and $\{S(k)\}$ is a sequence of
  i.i.d.\ Gaussian random variables} with zero mean and variance
  $\sigma_s^2.$ Thus, the intensity \an{evolves in time as follows:}
  \begin{equation}
    I(t) = \sum_{k=0}^{\infty} I(k) \mbox{rect}( t - m(k-1)),
    \label{eqn:intensity}
  \end{equation}
  \an{where} $\mbox{rect}(t)$ is the rectangular function \an{taking}
  value $1$ in the interval $[0,1]$ and zero elsewhere.

  \subsubsection*{Medium model}
  We model the warehouse \an{as a} rectangular region with known
  dimensions $l_1 \times l_2.$ Without loss of generality, we \an{let}
  the warehouse to be the region $D = [0,l_1] \times [0,l_2],$ 
  \an{and we denote the boundary of the warehouse by $\partial D$.}

  The medium is characterized by the diffusion coefficient of the gas,
  boundary conditions and initial conditions. We use $C(y,t;x)$ to
  denote the concentration at a point $y$ at time $t$ when the source
  is at $x.$ We make the following assumptions.
  \begin{enumerate}
  \item The diffusion coefficient of the gas is the same everywhere in
  the warehouse. We use $\nu$ to denote this value.
  \item The boundaries of the room are insluated, i.e., there is no
  leakage out of the room, i.e., $\frac{\partial C(s,\cdot;x)
  }{\partial t} = 0, \forall s \in \partial D.$
  \item At time $t = 0$ the concentration is $0$ everywhere in the room,
  i.e., $C(\cdot,0;x) = 0.$
  \end{enumerate}

  \subsubsection*{Observation model}
  Let $s_i$ be the location of the $i$-th sensor. We \an{assume that
  all} sensors sense at the beginning of each \an{time} slot, 
  i.e., at time $k.$
  Then
  \begin{equation}
     \label{eqn:measure}
     R_{i}(k;x) = C(s_i,k;x) + N_{i}(k),
   \end{equation}
   where $N_{i}(k)$ is \an{a zero mean i.i.d.\ measurement noise
   with known variance $\sigma_{n}^2$.}

   \subsubsection*{Transport model}
   We assume that the transport of the gas in the warehouse obeys the
   diffusion equation. Therefore,
   \begin{equation}
     \frac{\partial C(y,t;x)}{\partial t} = \nu \nabla^2 C(y,t;x) + I(t)
     \bar{\delta}(y - x),
     \label{eqn:pde}
   \end{equation}
   with the initial and boundary conditions
   \begin{eqnarray*}
     C(s,0;x) &= 0 & \mbox{for all $s \in D$,} \\ \frac{\partial
    C(s,t;x) }{\partial t} &= 0 & \mbox{for all $t\geq 0$ and $s \in
    \partial D$.}
  \end{eqnarray*}
  Here, $\nabla^2$ is the Laplacian differential operator and
  $\bar{\delta}$ is the Dirac delta function.

   \subsection{Problem statement and related literature}
   The medium characteristics \an{are} completely known, i.e., $l_1,$
   $l_2,$ and $\nu$ are known. The sensors' sampling duration and the
   measurement noise variance $\sigma_n^2$ are also known. Further,
   the variance \an{$\sigma^2_s$ of $S(k)$ is known}. The problem is
   to determine the location of the point source $x$ from the sensor
   measurements in a distributed and recursive manner. To solve the
   above problem we first show that, as a consequence of the
   assumptions that have been made, the sensor measurements follow a
   state-space model. We then use the IRPE algorithm developed in the
   previous sections to solve the problem.

   We next compare and contrast the models described above with the
   models used in literature. The point source model is a common model
   for diffusing sources and has been extensively used in localization
   studies \cite{Matthes05,Zhao07,Levinbook04,Alpay00}.  The random
   time-varying source intensity model is more realistic compared to
   the constant intensity \cite{Alpay00,Matthes05,Levinbook04} and
   instantaneous intensity models that are usually
   studied. Localization of sources with time-varying intensity have
   been studied in a centralized and non-recursive setting in
   \cite{Cannon98,Niliote01}. These studies consider a deterministic
   \an{evolution of the leak intensity and use a continuous
   observation model.} We are not aware of any paper that models the
   time-varying intensity as a random process. Most papers study that
   case when the medium is infinite or semi-infinite since the
   diffusion equation has a closed form solution in that case
   \cite{Levinbook04,Matthes05}.  The medium model assumed in this
   paper is more general. We also remark that we can extend the
   results to non-rectangular geometries by using the Galerkin
   approximation \cite{Sundhar07d}.

   While centralized recursive source localization has recieved much
   interest \cite{Matthes05,Piterbarg97,Niliote01,Levinbook04} there
   are very few papers that discuss a distributed solution. A
   recursive and distributed solution to the problem in a Bayesian
   setting is discussed in \cite{Zhao07}. A related paper is
   \cite{Rossi04} that deals with the problem of estimating the
   diffusion coefficient in a distributed and recursive manner.  We
   are not aware of any prior work that solves the source localization
   problem using a distributed and recursive approach in a
   non-Bayesian setting.

   \subsection{Approach}
   We show in Appendix~\ref{ssec:hs} that by using Green's technique
   to solve differential equations it is possible to obtain a
   state-space description for each sensor's observation process. We
   can then use the IRPE algorithm to estimate the iterates in a
   distributed and recursive manner.

   \subsection{Numerical results}
\begin{figure}[tb]
       \center{\includegraphics[scale = 0.262]{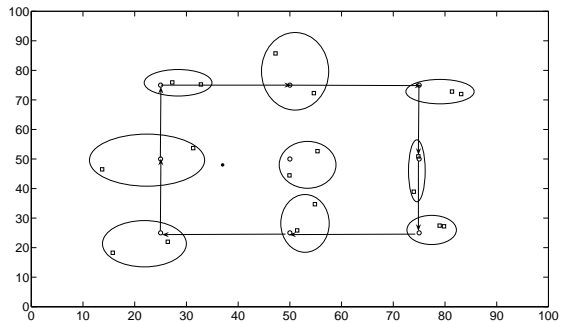}}
       \caption{A network of $27$ sensors. The circles denote the
       cluster heads and the squares denote the sensors. The source is
       represented by a dot. The arrows indicate the order in which
       the iterates are passed in the hybrid IRPE.}
       \label{Fig:network}
     \end{figure}

    We \an{use} $l_1 = l_2 = 100$ and diffusion coeffient
   \an{$\nu=~1.$} The actual location of the source is $x^* =
   (37,48).$ The initial intensity value is taken to be $100,$ $\rho$
   is fixed at $0.99$ and the variance of $S(k)$ is fixed at $10.$ A
   network of $27$ sensors \an{is} deployed. To ensure complete
   coverage of the sensing area, we first placed $9$ sensors on a grid
   and then randomly deployed $2$ sensors in the immediate
   neighborhood of each of the $9$ sensors. The network is shown in
   Fig.~\ref{Fig:network}.

   %\begin{center}
   %\end{center}   

   %\begin{center}
     \begin{figure}[tb]
       \center{\includegraphics[scale = 0.262]{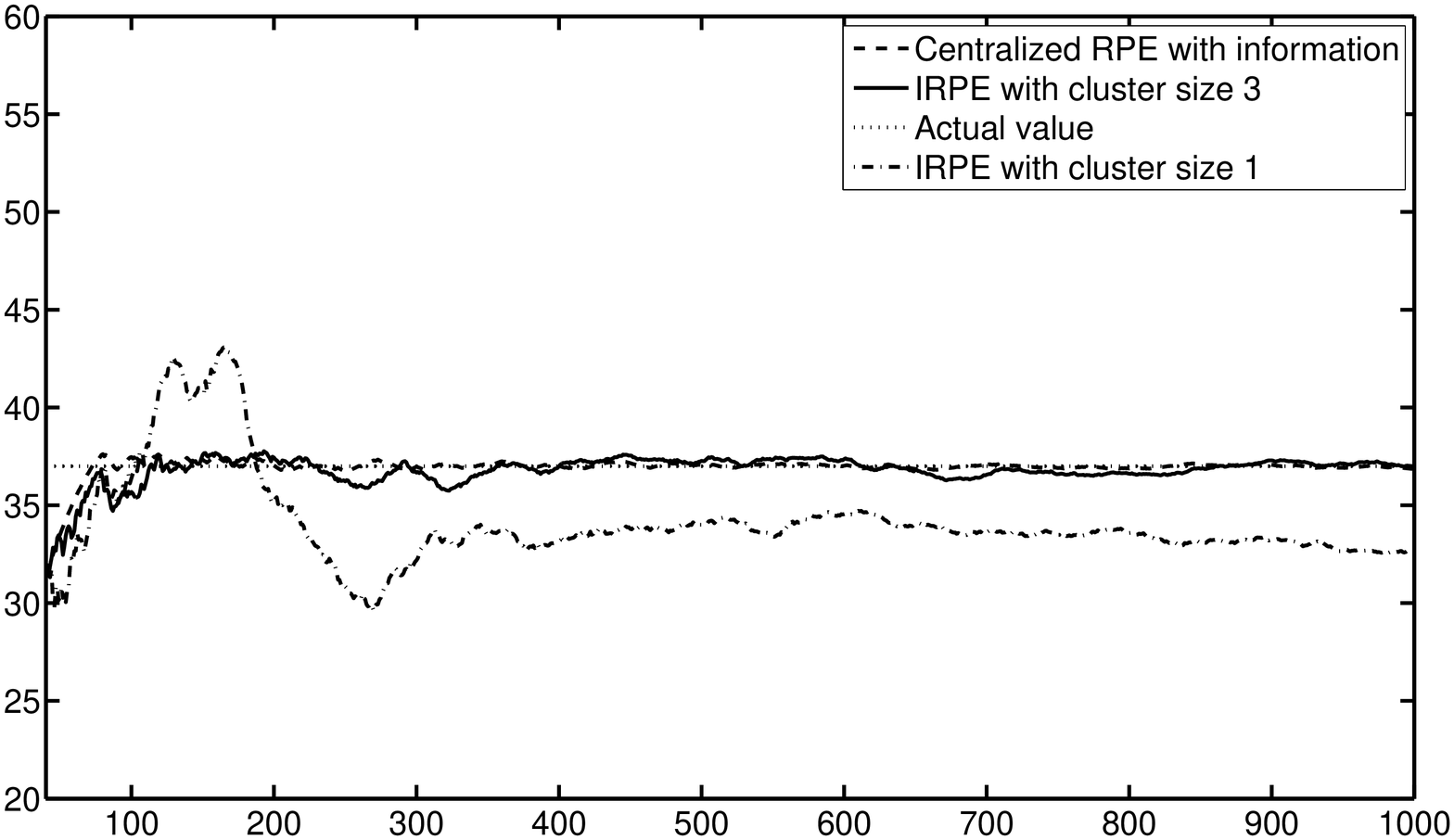}}
       \caption{Estimate of the $x$-coordinate generated by the standard
	 IRPE, hybrid IRPE and the centralized RPE. Observe that the
	 standard IRPE iterates get caught in a local minimum.}
       \label{Fig:xc}
     \end{figure}
   %\end{center}   
   %\begin{center}
     \begin{figure}[tb]
       \center{\includegraphics[scale = 0.262]{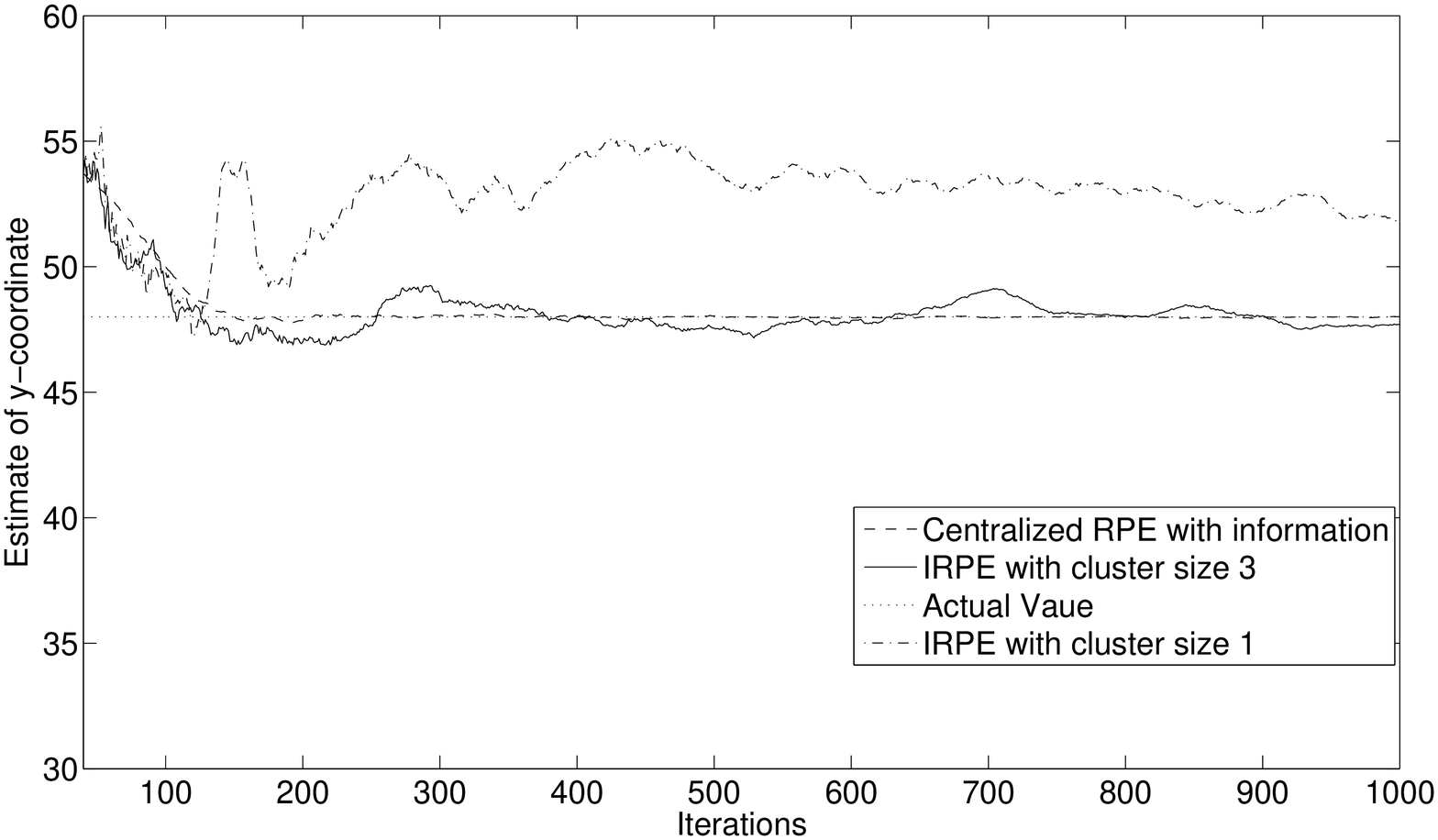}}
       \caption{Estimate of the $y$-coordinate generated by the standard
	 IRPE, hybrid IRPE and the centralized RPE. Observe that the
	 standard IRPE iterates get caught in a local minimum.}
       \label{Fig:yc}
     \end{figure}
   %\end{center}   

   The sampling interval \an{is} $10$ time units and the
   measurement noise variance \an{is set to} $0.1.$ \an{In deriving} the
   state-space representation, we \an{use} $\bar{n}_1 = \bar{n}_2 = 15.$ We
   performed three simulation experiments.
   \begin{enumerate}
   \item {\it Standard IRPE}: 
     1000 iterations of the IRPE algorithm \an{are} used
     to estimate the source location. As discussed, the IRPE algorithm
     does not use the information that the sensors observe the same
     under-lying process through different observation matrices $H_i.$
   \item {\it Hybrid IRPE}: The network \an{is} divided into $9$ clusters of
     size $3.$ The cluster heads \an{are} the sensors on the
     grid. At the beginning of each slot, \an{a} cluster head \an{collects}
     the measurements from the sensors in the cluster. 
     \an{To estimate the sensor location, 1000 iterations
     of IRPE are run between the cluster heads. 
     In} determining the predictor family
     for each cluster's observations the information that all the
     sensors in the cluster observe the same underlying process is
     used. But, in the inter-cluster processing through the IRPE
     algorithm this information is not used.
   \item \an{{\it Centralized RPE}: All sensors} immediately
     communicate their measurements to a fusion center. In this case
     the information is completely used. The fusion center runs 1000
     iterations.
   \end{enumerate}
   The results are plotted in Figs.\ \ref{Fig:xc}~and~\ref{Fig:yc}. As
   expected the centralized RPE performs the best. However, what is
   interesting to note is that the standard IRPE does not converge to
   the correct solution but is caught in a local minimum. We also
   observed this in other simulation runs. However, when the sensors
   are clustered the iterates converge to the correct location.

  \section{Conclusions}
  \label{sec:conclusions}
  Linear state-space models arise naturally, or as linear
  approximations to non-linear state-space models, in many
  applications. In an inference setting where the aim is to estimate a
  quantity of interest, it is quite natural for the state-space models
  to be parametrized by the unknown quantity of interest. In a control
  system setting where the aim is to control the process, it is common
  to use such incomplete state-space models as `grey-box' descriptions
  of the system that is to be controlled. Thus, the problem addressed
  in this paper is important in both of these settings. 

  In Section~\ref{sec:ex} we could only give a qualitative description
  of the tradeoff between centralized and distributed schemes. It is
  therefore of interest to find good bounds on the performance of the
  IRPE and RPE schemes that can be used to quantify the loss in
  performance. Also, to truly understand the performance of the
  algorithm in practical settings, we need to obtain convergence
  results when there are communication errors. Further, we have
  considered a simple class of networks where the topology is
  fixed. It is important to obtain an algorithm that is similar to the
  IRPE for networks with a random and time-varying
  topologies. Finally, as mentioned in Section~\ref{sec:ex}, the
  result and analysis extend easily to the case where there is an
  open-loop input to the system. An important extension is to obtain
  similar convergence results for some common classes of closed-loop
  inputs using the techniques discussed in Appendix~7.A of
  \cite{Ljung83}.

  \bibliographystyle{IEEEbib} 
  \bibliography{irpe}
  \appendix
  \subsection{Proof of Theorem~\ref{thm:IRPE}}
  \label{ssec:proof}
  We take the following approach to prove
  Theorem~\ref{thm:IRPE}. First, we consider a centralized system
  where the sensors immediately communicate all their measurements to
  a fusion center. For this system we use the RPE algorithm to
  generate a sequence of iterates, \an{and then we show that this
  iterate sequence is identical to the iterate sequence generated by
  the IRPE algorithm}. We complete the proof by proving that the
  iterates converge to a local minimum of $f(x).$

  \an{In what follows, we extensively use the notion} of block vectors
  and matrices. For positive integers $a$ and $b,$ let $\M_{a \times
  b}$ be the vector space of all real matrices of dimensions $a \times
  b.$ A block vector in $\M_{a \times b}$ is a vector whose elements
  are from $\M_{a \times b}.$ The length of a block vector is the
  number of block elements. In a similar manner, block matrices in
  $\M_{a \times b}$ are matrices where each element is itself a matrix
  from $\M_{a \times b}$.  While writing block matrices we will allow
  for a slight abuse of notation and use $0$ and $I$ to denote the
  zero and identity matrices, respectively. Their dimensions can be
  unambiguously fixed from the dimensions of the other blocks in the
  block matrix.  We will use $\bm{U}^{a}_{b},$ $b \leq m,$ to denote
  the unit block vector in $\M_{a \times a}$ of length $m,$ with the
  $b$-th block equal to the identity matrix in $\M_{a \times a}$ and
  all the other blocks equal to the zero matrix in $\M_{a \times a}.$

  We allow $i,j$ to take values in the set $\mathcal{I} =
  \{1,\ldots,m\}.$ We define $\delta[\cdot]$ as the Kronecker
  delta. Recall that the {\an dimension of the matrices
  $\Theta_i(k;x)$ is $q,$ the dimension of the measurement $r_{i}(k)$
  is $p,$ and the dimension of the parameter vector} $x$ is $d.$ Also,
  recall that for any random process $\{Y(k;x)\}$ that is parametrized
  by $x,$ $y(k)$ denotes the sample path of $Y(k;x^*).$

\remove{\todo{Check that all sections, subsections use capital letters
consistently}}
\subsubsection{State-space model for sensor observations}
  Without loss of generality, assume that each time slot \an{has
  duration of $m$} time units. Consider a hypothetical centralized scheme
  where at time $mk+j,$ sensor $j$ communicates $r_{j}(k+1)$ to the
  fusion center over a perfect delayless link. For $i \neq j,$ sensor
  $i$ communicates a predetermined constant value, \an{say $0$,} that
  does not convey any information about the value taken by the
  parameter $x.$

  \an{Denote the sequence communicated by a sensor $i$ by 
  $\{\bar{r}_i(mk+j)\}$, with}
  \begin{equation}
  \bar{r}_{i}(mk+j) = r_{i}(k+1) \delta[i-j].  \label{eqn:eob1}
  \end{equation}
  \an{Next, denote the observation sequence at the fusion center by 
  $\{\tilde{r}(mk+j)\}$, where}
  \begin{equation}
  \tilde{r}(mk+j) = \left[ \bar{r}_{1}(mk+j) \,\ldots\,
  \bar{r}_{m}(mk+j) \right]^T = \bm{U}^{p}_{j} r_j(k+1)
  . \label{eqn:eob2}
  \end{equation}
  
  \an{We now consider the problem of estimating $x$ from
  observation sequence $\{\tilde{r}(mk+j)\}.$
  We show that the random process $\tilde{R}(mk+j;x)$
  can be represented} as the output
  vector of a suitably defined state-space system. 
  \an{For this, we first use the relations in}
  (\ref{eqn:statespace}) to obtain the equations describing the
  evolution $\{\bar{R}_i(mk+j;x)\}.$ \an{Note that from (\ref{eqn:eob1}), 
  we have}
  \begin{equation}
  \bar{R}_i(mk+j;x) = R_i(k+1;x)\ \delta[i-j]. \label{eqn:rp}
  \end{equation}
  \an{Let $\bar{D}_{i}(x)$ be} the following $m \times m$ block
  matrix in $\M_{q\times q}$:
  \begin{align}
    \bar{D}_i(x) = \left[ \begin{array}{cccccc}
	0 & I & 0 &\cdot & \cdot & 0 \\
	0 & 0 & I &\cdot & \cdot & 0 \\
	\cdot & \cdot & \cdot & \cdot & \cdot & \cdot \\
	0 & 0 & 0 &\cdot & \cdot & I \\
	D_i(x) & 0 & 0 &\cdot & \cdot & 0 
      \end{array} \right]. \label{eqn:matd}
  \end{align}
  Observe that
  \begin{align}
    \bar{D}_i(x) \bm{U}^q_j = 
    \begin{cases}
      \bm{U}^q_{j-1} & \mbox{ when $j \neq 1$} \\
      \bm{U}^q_{m}\ D_i(x) & \mbox{ when $j =1$}.
    \end{cases} \label{eqn:key1}
  \end{align}
  Also, define $\bar{H}_i = H_i \,\left(\bm{U}^q_{1} \right)^T,$
  \an{and note} that
  \begin{align}
    \bar{H}_i \bm{U}^q_j = H_i \ \left(\bm{U}^q_{1} \right)^T
    \bm{U}^q_j = H_i \delta[j - 1].
     \label{eqn:key2}
  \end{align}
  Define $\bar{\Theta}_{i}(0;x) = \bm{U}^q_{i}\ \Theta_{i}(0;x),$ and
  \begin{align}
    &\bar{\Theta}_{i}(mk+j;x) =
    \begin{cases}
      \bm{U}^q_{i-j+1} \ \Theta_{i}(k+1;x) \ \ \ &\mbox{
	if $j \leq i$} \\ \bm{U}^q_{m+1-(j -i)} \
	\Theta_{i}(k+2;x) &\mbox{ if $j > i$,}
    \end{cases}  \label{eqn:defn1}\\
    &\bar{W}_{i}(mk+j;x) = \bm{U}^q_{m} \ W_{i}(k+1;x) \ \delta[i-j],
    \nonumber \\ &\bar{V}_{i}(mk+j) = V_{i}(k+1) \
    \delta[i-j]. \label{eqn:defn1b}
  \end{align}
  \an{The following is an illustration for $
   \bar{\Theta}_{i}(mk+j;x)$ with $i=3$ and $j=2,3,4$:} 
  \[
    \begin{array}{ccc}
      mk + 2 & mk + 3 & mk +4 \\
      \\
      \left[ 
	\begin{array}{c}
	  0 \\
	  \Theta_{3}(k+1;x)\\
	  0 \\
	  \vdots \\ 
	  0 \\
	  0
	\end{array}
	\right]
      &
      \left[ 
	\begin{array}{c}
	  \Theta_{3}(k+1;x) \\
	  0\\
	  0\\
	  \vdots \\ 
	  0 \\
	  0
	\end{array}
	\right]
      &
      \left[ 
	\begin{array}{c}
	  0 \\
	  0\\
	  0\\
	  \vdots \\ 
	  0 \\
	  \Theta_{3}(k+2;x)
	\end{array}
	\right].
    \end{array}
    \]
    \begin{claim}
    \label{claim1}
    \an{For all $n\ge0,$ we have}
    \begin{align}
      \bar{\Theta}_i(n+1;x) &= \bar{D}_i(x) \bar{\Theta}_{i}(n;x) +
      \bar{W}_{i}(n;x), \label{eqn:newss1} \\ \bar{R}_{i}(n+1;x) &=
      \bar{H}_i \bar{\Theta}_{i}(n+1;x) +
      \bar{V}_{i}(n+1). \label{eqn:newstatespace}
    \end{align}
  \end{claim}
  \begin{proof}
    Let $n = mk+j.$ Substitute for $\bar{\Theta}_{i}(n;x)$ and
    $\bar{W}_{i}(n;x)$ from (\ref{eqn:defn1}) in the \an{right hand side
    (RHS)} of
    (\ref{eqn:newss1}). For $i < j,$ from (\ref{eqn:key1}) \an{we obtain}
    \begin{align*}
      \mbox{RHS of
    (\ref{eqn:newss1})} &= \bar{D}_i(x) \bm{U}^q_{i-j+1} \ \Theta_{i}(k+1;x)
     + 0 \\& =\bm{U}^q_{i-j} \Theta_{i}(k+1;x) \\& = \bm{U}^q_{i-(j+1)
     - 1} \Theta_{i}(k+1;x) \\&= \bar{\Theta}_i(mk+j+1;x).
    \end{align*}
    For $i=j,$  using  (\ref{eqn:key1}) and  (\ref{eqn:statespace}), 
    \an{we obtain}
    \begin{align*}
      \mbox{RHS of 
    (\ref{eqn:newss1})} &= \bar{D}_i(x) \bm{U}^q_{1} \ \Theta_{i}(k+1;x) +
      \bm{U}^q_{m} \ W_{i}(k+1;x) 
      \\&=\bm{U}^q_{m} D_i(x) \Theta_{i}(k+1;x) + \bm{U}^q_{m} \ W_{i}(k+1;x)
      \\
      &=\bm{U}^q_{m} (  D_i(x) \Theta_{i}(k+1;x) + W_{i}(k+1;x) ) \\
      &= \bm{U}^q_{m} \Theta_{i}(k+2;x) \\&= \bar{\Theta}_i(mk+j+1;x).
    \end{align*}
    Finally, when $i < j,$ from (\ref{eqn:key1}) \an{we have}
    \begin{align*}
      \mbox{RHS of
    (\ref{eqn:newss1})} &= \bar{D}_i(x) \bm{U}^q_{m-(j-i)+1} \
      \Theta_{i}(k+2;x) + 0 \\ &=\bm{U}^q_{m - (j-i)}
      \Theta_{i}(k+2;x) \\&= \bm{U}^q_{m - (j+1-i) - 1}
      \Theta_{i}(k+2;x) \\&=\bar{\Theta}_i(mk+j+1;x),
    \end{align*}
    \an{thus, completing the proof of 
    the relation in} (\ref{eqn:newss1}). 

    \an{We next prove the relation in (\ref{eqn:newstatespace}). 
    At first, we consider the case when $j\ne m$ and show that
    the following relation holds:} 
    \[
    \bar{H}_i \bar{\Theta}_{i}(mk+j+1;x) + \bar{V}_{i}(mk+j+1) =
    R_{i}(k+1) \delta[i-j-1].
    \]
    \an{Let $i \neq j+1,$ and note that from (\ref{eqn:defn1b}) we have}
    $\bar{V}_{i}(n+1) = 0.$ \an{Furthermore}, from the definition of
    \an{$\bar{\Theta}_i(n;x)$} in (\ref{eqn:defn1}) \an{we obtain}
    {\small
      \[
      \bar{H}_i \bar{\Theta}_i(mk+(j+1);x) = \begin{cases}
	\bm{U}^q_{i-j} \Theta_i(k+1;x) & \mbox{ $i > j+1$} \\
	\bm{U}^q_{m-(j-i)} \Theta_i(k+2;x) & \mbox{ $i < j+1$}
    \end{cases}
    \]
    }
    \an{Using} the expression in (\ref{eqn:key2}), \an{we} can immediately
    verify that $\bar{H}_i \bar{\Theta}_i(mk+(j+1);x) = 0.$ Therefore,
    \an{we have $\bar{R}_i(n+1;x)=0$ for $i \neq j+1.$}  
    
    When $i = j+1,$ $\bar{V}_{i}(mk+j+1) = V_{j+1}(k+1)$ and
    \[
    \bar{H}_{j+1} \bar{\Theta}_i(mk+j+1;x) = \bar{H}_{j+1}
    \bm{U}^q_{1} = H_{j+1} \Theta_{j+1}(k+1;x).
    \]
    Therefore, from (\ref{eqn:statespace}) we see that
    \an{$\bar{R}_i(mk+j+1;x)=R_i(k+1;x)$ for $i = j+1,$ thus,
    concluding the proof of relation (\ref{eqn:newstatespace}) for
    $j\ne m.$}
 
    When $j=m,$ \an{by using arguments similar to that of the
    preceding case $j \ne m,$ we can show} that
    \[
    \bar{H}_i \bar{\Theta}_{i}(mk+j+1;x) + \bar{V}_{i}(mk+j+1) =
    R_{1}(k+2) \delta[i-1],
    \]
    \an{thus, completing} the proof. 
  \end{proof}
  
  \an{Now, by combining the equations in 
  (\ref{eqn:newss1})--(\ref{eqn:newstatespace})
  for $i \in \mathcal{I},$ we provide evolution equations for
  $\{\tilde{R}(n;x)\}.$}
  Define
  \begin{align}
    &\tilde{F}(x) = \mbox{diag}\left( \bar{F}_1(x), \ldots,
    \bar{F}_m(x) \right), \nonumber \\ &\tilde{H}(x) =
    \mbox{diag}\left( \bar{H}_1(x), \ldots, \bar{H}_m(x) \right),
    \nonumber \\ &\tilde{\Theta}(n;x) = \left[
      \begin{array}{c}
      \bar{\Theta}_1(n;x) \\ \vdots\\ \bar{\Theta}_m(n;x) \end{array}\right],
  \ \ 
    \tilde{W}(n;x) =
    \left[
      \begin{array}{c}
      \bar{W}_1(n;x) \\ \vdots \\ \bar{W}_m(n;x) \end{array} \right],
    \ \ \tilde{V}(n;x) = \left[ 
       \begin{array}{c}
      \bar{V}_1(n;x) \\ \vdots \\ \bar{V}_m(n;x) \end{array} 
    \right].  \label{eqn:def}
  \end{align}
  \an{Using the relations in} (\ref{eqn:newss1})~and~(\ref{eqn:newstatespace}),
  we can write
  \begin{align}
    \tilde{\Theta}(n+1;x) &= \tilde{D}(x) \tilde{\Theta}(n;x) +
    \tilde{W}(n;x), \label{eqn:nss2} \\ \tilde{R}(n+1;x) &=
    \tilde{H} \tilde{\Theta}_{i}(n+1;x) +
    \tilde{V}(n+1). \label{eqn:nss}
  \end{align}

  To apply the RPE we need to determine a predictor family for
  $\tilde{R}(n;x^*)$ that is parametrized by $x$ and is
  asymptotically optimal at $x = x^*.$ We do \an{this in the next
  section}.

  \subsubsection{Time-Invariant Kalman Predictor for  Centralized System}
  Let us first obtain the time-invariant Kalman predictor for
  $\bar{R}_i(n;x)$.  Fix $n = mk+j$ and define
  \begin{align}
    \bar{\phi}_{i,n}(x;\bar{r}_i^{n-1}) &=
    \begin{cases}
      \bm{U}^q_{i-j+1} \ \phi_{i,k+1}(x;r_i^{k}) \ \ \ &\mbox{
	if $j \leq i$} \\ \bm{U}^q_{m+1-(j -i)} \
	\phi_{i,k+2}(x;r_i^{k+1}) &\mbox{ if $j > i$,}
    \end{cases} \label{eqn:if2} \\
     &\bar{g}_{i,n}(x;\bar{r}_{i}^{n-1}) = g_{i,k+1}(x;r_{i}^{k}) \
     \delta[j-i].  \label{eqn:if2b} 
  \end{align}
  Note that $\phi_{i,k+1}(x^*;r_i^{k})$ and
  $g_{i,k+1}(x^*;r_{i}^{k}),$ are the time-invariant Kalman predictors
  for $\Theta_{i}(k+1;x^*)$ and $R_{i}(k+1;x^*),$
  respectively. Therefore, from (\ref{eqn:defn1}) we can conclude that
  $\bar{\phi}_{i,n}(x^*;\bar{r}_i^{n-1})),$
  resp. $\bar{g}_{i,n}(x^*;\bar{r}_{i}^{n-1}),$ is asymptotically
  optimal for $ \bar{\Theta}_{i}(n;x^*),$ resp. $\bar{R}_i(n;x^*)$.

  Define
  \begin{align}
  \bar{G}_{i}(x) = \bm{U}^{p}_{m}\ G_i(x), \label{eqn:bG}
  \end{align}
  and $\bar{F}_i(x) = \bar{D}_i(x) - \bar{G}_{i}(x) \bar{H}_{i}.$ The
  matrix $\bar{F}_i(x)$ will have the same form as $\bar{D}_i(x)$ in
  (\ref{eqn:matd}) but with $D_i(x)$ replaced by $F_i(x).$ 
  \an{Similar to Claim~\ref{claim1}, we can show} that
  \begin{align}
    \bar{\phi}_{i,n+1}(x;\bar{r}_{i}^{n}) &= \bar{F}_{i}(x)
    \bar{\phi}_{i,n}(x;\bar{r}_{i}^{n-1}) + \bar{G}_{i}(x)
    \bar{r}_{i}(n), \nonumber \\ \bar{g}_{i,n+1}(x;\bar{r}_{i}^{n}) &=
    \bar{H}_{i}(x) \bar{\phi}_{i,n}(x;\bar{r}_{i}^{n}).
    \label{eqn:newdet}
  \end{align}

  We can immediately obtain a predictor family for
  $\tilde{\Theta}(n;x^*)$ and $\{\tilde{R}_{n}(x^*)\}$ that is
  asymptotically optimal at $x = x^*$ as follows:
  \begin{align*}
    \tilde{\phi}_n(x;\tilde{r}^{n-1}) &=
    \left[\bar{\phi}_{1,n}(x;\bar{r}_{1}^{n-1}) \ldots
    \bar{\phi}_{m,n}(x;\bar{r}_{m}^{n-1}) \right]^T, \\
    \tilde{g}_n(x;\tilde{r}^{n-1}) &=
    \left[\bar{g}_{1,n}(x;\bar{r}_{1}^{n-1}) \ldots
    \bar{g}_{m,n}(x;\bar{r}_{m}^{n-1}) \right]^T.
  \end{align*}
  \an{Furthermore}, from (\ref{eqn:newdet}) one can verify that
  \begin{align}
    \tilde{\phi}_{n+1}(x;\tilde{r}^{n}) &= \tilde{F}(x)
    \tilde{\phi}_{n}(x;\tilde{r}^{n-1}) + \tilde{G}(x)
    \tilde{r}(n), \nonumber \\ \tilde{g}_{n+1}(x;\tilde{r}^{n}) &=
    \tilde{H} \phi_{n+1}(x;\tilde{r}^{n}), \label{eqn:ndm}
  \end{align}
  \an{where} 
  \begin{equation}
  \tilde{G}(x) =
  \mbox{diag}\left(\bar{G}_1(x),\ldots,\bar{G}_m(x)\right). \label{eqn:tG}
  \end{equation}
  
  \subsubsection{RPE Algorithm for Centralized System}
  \an{Here, we use} the RPE algorithm to estimate $x$ from
  $\{\tilde{r}(n)\}.$ As we mentioned earlier, $\tilde{r}(n)$ 
  \an{contans the same information as $\hat{r}(n)$
  about the true value of $x$}.  Define \an{for $\ell=1,\ldots,d,$}
  \[
    \nabla^{(\ell)} \tilde{F}(x) = \frac{ \partial \tilde{F}(x) }{\partial
      x^{(\ell)}}, \qquad \nabla^{(\ell)} G(x) = \frac{ \partial \tilde{G}(x)
      }{\partial x^{(\ell)}}.
  \]
  \an{We define the iterates $\{\tilde{x}_{n}\}$ as follows:}
  \begin{eqnarray}
    \left[
      \begin{array}{c}
		\tilde{h}_{n+1}\\ 
	        \tilde{\xi}^{(\ell)}_{n+1}
      \end{array}
      \right] &=& \left[ \begin{array}{cc}
	\tilde{H} & 0 \\0 &  \tilde{H} 
	\end{array}\right] 
    \left[ \begin{array}{c} 
		\tilde{\psi}_{n+1} \\ 
		\tilde{\chi}^{(\ell)}_{n+1}
      \end{array}  \right], \label{eqn:CRPE}  \\
    \tilde{\epsilon}_{n+1} &=& \tilde{r}(n+1) - \tilde{h}_{n+1}, 
     \label{eqn:CRPE1} \\ 
    \underline{\tilde{x}}^{(\ell)}_{n+1} &=& \tilde{x}^{(\ell)}_{n} -
    \tilde{\alpha}_{n+1}\ \left(\tilde{\xi}^{(\ell)}_{n+1}\right)^T \, 
    \tilde{\epsilon}_{n+1},
    \label{eqn:CRPE2} \\ \underline{\tilde{x}}_{n+1} &=& \left[
    \underline{\tilde{x}}^{(1)}_{n+1} \,\ldots \, 
    \underline{\tilde{x}}^{(d)}_{n+1} \right]^T, \label{eqn:CRPE3} \\
    \tilde{x}_{n+1} &=& \mathcal{P}_X\left[
    \underline{\tilde{x}}_{n+1} \right], \label{eqn:CRPE10} \\ \left[
    \begin{array}{c} \tilde{\psi}_{n+2} \\ \tilde{\chi}^{(\ell)}_{n+2}
      \end{array}  \right]
    &=& \left[ \begin{array}{cc} \tilde{F}(\tilde{x}_{n+1}) & 0 \\
	  \nabla^{(\ell)} \tilde{F}(\tilde{x}_{n+1}) &
	  \tilde{F}(\tilde{x}_{n+1})
      \end{array}
      \right]  \left[ \begin{array}{c}
    \tilde{\psi}_{n+1} \\ \tilde{\chi}^{(\ell)}_{n+1}
      \end{array}  \right] + \left[ \begin{array}{c}
	\tilde{G}(\tilde{x}_{n+1}) \\
	\nabla^{(\ell)} G(\tilde{x}_{n+1})
	 \end{array} 
      \right] \tilde{r}(n+1). \label{eqn:CRPE4}
  \end{eqnarray}
  Here, \an{$\alpha(n)=\alpha_{k+1}$ for $n=mk+j$ for $j
  =1,\ldots,m-1.$ Next, we assign} the initial values for the
  recursion.  Recall that the IRPE algorithm in (\ref{eqn:IRPE4}) is
  initialized with the values $\psi_{i,1} = \psi_{i,s},$
  $\xi^{(\ell)}_{i,1} = \xi^{(\ell)}_{i,s}$ for all $i$ \an{and $\ell$,}
  and $x_{0} = x_s.$ We \an{let $\tilde{x}_{0} = x_s,$ and}
  \begin{align}
   \tilde{\psi}_{0} = \left[
      \begin{array}{c}
	\bar{\psi}_{1,s} \\ \vdots\\ \bar{\psi}_{m,s}
	\end{array}\right], \ \ \ 
   \tilde{\xi}^{(\ell)}_{0} = \left[
      \begin{array}{c}
	\bar{\xi}^{(\ell)}_{1,s} \\ \vdots\\ \bar{\xi}^{(\ell)}_{m,s}
	\end{array}\right],
  \end{align}
  \an{where} $\bar{\psi}_{i,s} = \bm{U}^q_{i} \psi_{i,s}$ and
  $\bar{\xi}^{(\ell)}_{i,s} = \bm{U}^q_{i} \xi^{(\ell)}_{i,s}$
  \an{for all $i$ and $l.$}

  \subsubsection{Rest of the proof}
  \an{Finally, here we}  show that $\tilde{x}_n = z_{j,k+1}.$ Recall that
  $\psi_{i,k}$ and $\chi^{(\ell)}_{i,k}$ \an{are} generated in the IRPE
  algorithm \an{(\ref{eqn:IRPE1})--(\ref{eqn:IRPE4})}. 
  Define for $l = 1,\ldots,d,$
   \begin{align}
    \bar{\psi}_{i,n} &=
    \begin{cases}
      \bm{U}^q_{i-j+1} \ \psi_{i,k+1} \ \ \ &\mbox{
	if $j \leq i$} \\ \bm{U}^q_{m+1-(j -i)} \
	\psi_{i,k+2} &\mbox{ if $j > i$,}
    \end{cases} \label{eqn:ee1} \\
    \bar{\chi}^{(\ell)}_{i,n} &=
    \begin{cases}
      \bm{U}^q_{i-j+1} \ \chi^{(\ell)}_{i,k+1} \ \ \ &\mbox{
	if $j \leq i$} \\ \bm{U}^q_{m+1-(j -i)} \
	\chi^{(\ell)}_{i,k+2} &\mbox{ if $j > i$.}
    \end{cases} \label{eqn:ee2}
   \end{align}
   We next establish the following lemma.  Once we prove this lemma we
   can use induction to \an{show} that the iterates generated by the
   IRPE \an{algorithm are} the same as those generated by the
   centralized scheme.  \emph{
    \begin{lemma}
      \label{lemma1}
      \an{Let $n = mk+j.$ If $\tilde{x}_{n} = z_{j,k+1}$ and} 
      \begin{align}
	\tilde{\psi}_{n+1} &= \left[\bar{\psi}_{1,n+1} \,\ldots \,
	\bar{\psi}_{m,n+1} \right]^T, \label{l1} \\
	\tilde{\chi}^{(\ell)}_{n+1} &=
	\left[\bar{\chi}^{(\ell)}_{1,n+1} \,\ldots \,
	\bar{\chi}^{(\ell)}_{m,n+1} \right]^T\qquad \an{\hbox{for }
	\ell=1,\ldots,d,} \label{l2}
      \end{align}
      then $\tilde{x}_{n} = z_{j+1,k+1},$ \an{and}
      \begin{align*}
	\tilde{\psi}_{n+2} &= \left[\bar{\psi}_{1,n+2} \,\ldots \,
	\bar{\psi}_{m,n+2} \right]^T, \\ \tilde{\chi}^{(\ell)}_{n+2}
	&= \left[\bar{\chi}^{(\ell)}_{1,n+2} \,\ldots \,
	\bar{\chi}^{(\ell)}_{m,n+2} \right]^T\qquad \an{\hbox{for }
	\ell=1,\ldots,d.}
      \end{align*}
    \end{lemma}
  }
  \begin{proof}
    \an{For a block vector $A,$ let $A^{(i)}$ denote} its $i$-th
    block.  Substituting for $\tilde{\psi}_{mk+j+1}$ from (\ref{l1})
    in (\ref{eqn:CRPE}), and noting from (\ref{eqn:def}) that
    $\tilde{H}$ is a block diagonal matrix with the $(i,i)$-th block
    equal to $\bar{H}_i$, we can see that
    \begin{align*}
      \tilde{h}_{mk+j+1}^{(i)} &= (\tilde{H} \tilde{\psi}_{mk+j+1}
                           )^{(i)} = \bar{H}_i \bar{\psi}_{i,mk+j+1}.
    \end{align*}
    Using the definition of $\bar{\psi}_{i,mk+j+1}$ from (\ref{eqn:ee1})
    and noting from (\ref{eqn:key2}) that $\bar{H}_i \bm{U}^q_{j} =
    H_i \delta[j-1]$, we \an{obtain}
    \begin{align*}
      \tilde{h}^{(i)}_{mk+(j+1)} &= \bar{H}_i \bar{\psi}_{i,mk+(j+1)}
      \\&=
      \begin{cases}
	\bar{H}_i \an{U}^{q}_{i - (j+1) -1} \psi_{i,k+1} & \mbox{ if $i >
	j+1$}\\ \bar{H}_{i} \an{U}^{q}_1 \psi_{i,k+1} & \mbox{ if $i =
	j+1$} \\ \bar{H}_{i} \an{U}^{q}_{m - (j+1-i)+1} \psi_{i,k+1} &
	\mbox{ if $i < j+1$}
      \end{cases}
      \\&= 
      \begin{cases}
	0 & \mbox{ if $i \neq j+1$}\\
	H_{j+1} \psi_{j+1,k+1}  & \mbox{ if $i = j+1$}. 
      \end{cases}
    \end{align*}
    Using (\ref{eqn:IRPE1}) we replace $H_{j+1} \psi_{j+1,k+1}$ by
    $h_{j+1,k+1}$ and write
    \begin{align*}
      \tilde{h}^{(i)}_{mk+(j+1)} = h_{i,k+1} \delta[i-j-1].
    \end{align*}
    Therefore, $\tilde{h}_{mk+j+1} = \bm{U}^{\an{p}}_{j+1}\ h_{j+1,k+1}.$ 
    \an{Similarly, we can see} that \an{for all $\ell$,}
    \begin{align}
    \tilde{\xi}^{(\ell)}_{mk+j+1} = \bm{U}^p_{j+1} \ 
    \xi^{(\ell)}_{j+1,k+1}. \label{eqn:k1}
    \end{align}
    Substituting in (\ref{eqn:CRPE1}) for $\tilde{h}_{mk+j+1}$ from
    above and for $\tilde{r}_{mk+j+1}$ from (\ref{eqn:eob2}) we get
    \begin{align*}
      \tilde{\epsilon}_{mk+j+1} = \bm{U}^p_{j+1} (h_{j+1,k+1} -
      r_{j+1}(k+1) ).
    \end{align*}
    \an{Observe that $\epsilon_{j+1,k+1} =
    h_{j+1,k+1} - r_{j+1}(k+1)$ from (\ref{eqn:IRPE2}), so that}
    \begin{align}
      \tilde{\epsilon}_{mk+j+1} = \bm{U}^p_{j+1} \epsilon_{j+1,k+1}.
      \label{eqn:k2}
    \end{align}

    Since $\tilde{x}_{mk+j} = z_{j,k+1}$ it follows that
    $\tilde{x}^{\an{(l)}}_{mk+j} = z^{(\ell)}_{j,k+1}.$ Substituting from
    (\ref{eqn:k1}) and (\ref{eqn:k2}) in (\ref{eqn:CRPE2}) we get
    \an{for all $\ell$,}
    {\small
     \begin{align*}
       \underline{\tilde{x}}^{(l)}_{mk+j+1} &= z^{(\ell)}_{j,k+1} -
       \alpha_{k+1} \left(\bm{U}^p_{j+1} \xi^{(\ell)}_{j+1,k+1} \right)^T
       \bm{U}^p_{j+1} \epsilon_{j+1,k+1} \\ &= z^{(\ell)}_{j,k+1} -
       \alpha_{k+1} \left(\xi^{(\ell)}_{j+1,k+1}\right)^T
       \epsilon_{j+1,k+1} \\ &= \underline{z}^{(\ell)}_{j+1,k+1}.
     \end{align*}
     } The last step follows from (\ref{eqn:IRPE3}). Therefore, from
     (\ref{eqn:CRPE3}) and (\ref{eqn:IRPE10}) we can conclude that
     $\underline{\tilde{x}}_{mk+j+1} = \underline{z}_{j+1,k+1}$ and
     from (\ref{eqn:CRPE10}) and (\ref{eqn:IRPE11}) that
     $\tilde{x}_{mk+j+1} = z_{j+1,k+1}.$ This completes the first part
     of the proof.

     Let us next consider the case when $j\in \{1,\ldots,m-1\}.$ In
     (\ref{eqn:CRPE4}) let us replace $\tilde{x}_{n+1}$ with
     $z_{j+1,k+1}.$ Note from (\ref{eqn:def}), respectively
     (\ref{eqn:tG}), that $\tilde{F}(z_{j+1,k+1}),$ respectively
     $\tilde{G}(z_{j+1,k+1}),$ is a block diagonal matrix with
     $(i,i)$-th block equal to $\bar{F}_{i}(z_{j+1,k+1}),$
     respectively $\bar{G}_{i}(z_{j+1,k+1}).$ Substituting for
     $\tilde{\psi}_{n+1}$ from (\ref{l1}) and $\tilde{r}(n+1)$ from
     (\ref{eqn:eob2}) in (\ref{eqn:CRPE4}) we can write
     \begin{align*}
       \tilde{\psi}^{(i)}_{mk+j+2} =& \bar{F}_{i}(z_{j+1,k+1})
       \bar{\psi}_{i,mk+j+1} + \bar{G}_{i}(z_{j+1,k+1})
       \bar{r}_{i}(mk+j+1).
     \end{align*}
     Let us substitute for $\bar{G}_{i}(z_{j+1,k+1})$ from
     (\ref{eqn:tG}), for $\bar{\psi}_{i,mk+j+1}$ from (\ref{eqn:ee1})
     and for $\bar{r}_{i}(mk+j+1)$ from (\ref{eqn:eob1}). Using
     (\ref{eqn:key1}) we get for $i> j+1,$
     \begin{align*}
	 \tilde{\psi}^{(i)}_{mk+j+2} &= \bar{F}_{i}(z_{i,k+1})
	 \bm{U}^q_{i-j} \psi_{i,k+1} + 0 \\ 
	 &= \bm{U}^q_{i-j-1} \psi_{i,k+1} \\&= \bar{\psi}_{i,mk+j+2}.
     \end{align*}
     When $i = j+1,$ from (\ref{eqn:IRPE4}) \an{we obtain}
     \begin{align*}
	 \tilde{\psi}^{(i)}_{mk+j+2} &= \bar{F}_{i}(z_{i,k+1})
	 \bm{U}^q_{1} \psi_{j+1,k+1} + \bar{G}_{i}(z_{j+1,k+1})
	 \bar{r}_{i}(n+1) \\ &= \bm{U}^q_{m} F_i(z_{i,k+1})
	 \psi_{i,k+1} + \bm{U}^q_{m} G_i(z_{i,k+1}) r_{i}(k+1) \\ &=
	 \bm{U}^q_{m} \psi_{i,k+2} 
         \\ &= \bar{\psi}_{i,mk+j+2}.
     \end{align*}
     When $i < j+1,$ \an{we have}
     \begin{align*}
	 \tilde{\psi}^{(i)}_{mk+j+2} &= \bar{F}_{i}(z_{i,k+1})
	 \bm{U}^q_{m - (j-i)} \psi_{i,k+2} + 0 \\ 
	 &= \bm{U}^q_{m -(j-i) - 1} \psi_{i,k+2} \\ 
         &= \bar{\psi}_{i,mk+j+2}.
     \end{align*}
     Equation (\ref{l1}) can be  \an{shown similarly} for the 
     case $j = m.$ The proof of relation (\ref{l2}) is very similar to
     \an{that of relation (\ref{l1}) and therefore, it is omitted}.
  \end{proof}
    
  Observe that the initial values for the RPE algorithm in
  (\ref{eqn:CRPE4}) and the IRPE algorithm in (\ref{eqn:IRPE4}) are
  choosen such that  $\tilde{x}_{0} = z_{0,1},$ \an{and}
  \begin{align*}
    \tilde{\psi}_{1} &= \left[\bar{\psi}_{1,1} \,\ldots \,
      \bar{\psi}_{m,1} \right]^T, \\
    \tilde{\chi}^{(\ell)}_{1} &= \left[\bar{\chi}^{(\ell)}_{1,1} \,\ldots \,
      \bar{\chi}^{(\ell)}_{m,1} \right]^T\quad
    \an{\hbox{for all $\ell$}.}
  \end{align*}
  \an{By using Lemma~\ref{lemma1} and the induction on $k$}, 
  we can conclude that $\tilde{x}_{mk+i} = z_{i,k+1}$ 
  for all $k\ge1$
  and $i \in \mathcal{I}$. To complete
  the proof, we only need to show that the  
  \an{sequence $\{\tilde{x}_n\}$ converges} to a
  minimum of $f(x),$ \an{which is done in the following}.\emph{
  \begin{lemma}
    The sequence $\{\tilde{x}_n\}$ \an{generated by
    (\ref{eqn:CRPE})--(\ref{eqn:CRPE4})} converges to a local minimum
    of the function $f(x),$ defined in (\ref{eqn:costfn}), over the
    set $X$ w.p.1.
  \end{lemma}
  }
  \begin{proof}
    \an{By the assumptions of Theorem~\ref{thm:IRPE}, it follows that
    the} conditions of Theorem~\ref{thm:rpe} \an{are satisfied}.
    Therefore, \an{by Theorem~\ref{thm:rpe}}, the iterates \an{$\tilde
    x_n$ converge to a local} minimum over the set $X$ of the
    following function
    \begin{align*}
      \tilde{f}(x) &=\lim_{N \to \infty} \frac{1}{N} \sum_{n=1}^{N}
      \EXP{\left\| \tilde{R}(n;x^*) - \tilde{g}_n(x, \tilde{R}^n(x^*))
      \right\|^2} \\ &=\lim_{N \to \infty} \frac{1}{N} \sum_{n=1}^{N}
      \sum_{i=1}^{m} \EXP{\left\| \bar{R}_i(n;x^*) - \bar{g}_{i,n}(x,
      \bar{R}_i^n(x^*)) \right\|^2} \\ &= \lim_{N \to \infty}
      \frac{1}{N} \sum_{n=1}^{N} \sum_{i=1}^{m} \EXP{\left\|
      R_i(n;x^*) - g_{i,n}(x;R_{\an{i}}^n(x^*)) \right\|^2} \\&=
      f(x).
    \end{align*}
  \end{proof}

  \subsection{State-space Model for the Sensor Measurements}
  \label{ssec:hs}
  A standard technique to solve partial differential equations with
  boundary and initial conditions is to use Green's function. For the
  equation in (\ref{eqn:pde}) the solution can be written as follows
  \[
  C(y,t;x) = \int_{0}^{t} \int_{D} I(\tau) \bar{\delta}(z - x) G_f(y,z,t
  - \tau) dz \ d\tau.
  \]
  Here, $\tau$ and $z = (z_1,z_2)$ are parameters of integration, and
  $G_f$ is the following Green's function
    \begin{align*}
      G_f(y,z,t) &= \frac{1}{l_1 l_2} + \sum_{n_1,n_2=1}^{\infty} \
      \prod_{i=1}^{2} \frac{2}{l_i} \exp\left( - \frac{ \nu n_i^2
      \pi^2 t}{l_i^2} \right) \cos\left( \frac{n_i \pi
      y_i}{l_i}\right) \cos\left( \frac{n_i \pi z_i}{l_i} \right).
    \end{align*}
    Evaluating $C(y,t;x),$ we obtain
    \begin{align}
       &C(y,t;x) = \frac{1}{l_1 l_2} \int_{0}^{t} I(\tau)\ d\tau +
       \sum_{n_1=1}^{\infty} \sum_{n_2=1}^{\infty} \prod_{i=1}^{2}
       \left( \frac{2}{l_i} \cos\left( \frac{n_i \pi y_i}{l_i}\right)
       \cos\left( \frac{n_i \pi x_i}{l_i} \right) \right) \left(
       \int_{0}^{t} I(\tau) \beta_{n_1,n_2}^{t - \tau} \ d\tau
       \right), \label{eqn:Gf}
    \end{align}
    \an{where}
   \[
   \beta_{n_1,n_2} = \exp\left( - \nu \pi^2 \sum_{i=1}^{2} \frac{
     n_i^2 }{l_i^2} \right).
  \]

   To get a convenient approximation we will use a sufficiently large,
   but fixed number of terms in the convergent infinite series in
   (\ref{eqn:Gf}).  We will let $n_i,$ $i=1,2,$ vary from $1$ to
   $\bar{n}_i,$ \an{where the integers $\bar{n}_i>0$ are 
   chosen large enough to provide}
   a sufficiently good approximation. Therefore,
    \begin{align}
       & C(y,t;x) \simeq \frac{1}{l_1 l_2} \int_{0}^{t} I(\tau)\ d\tau
       + \sum_{n_1=1}^{\bar{n}_1} \sum_{n_2=1}^{\bar{n}_2}
       \prod_{i=1}^{2} \left( \frac{2}{l_i} \cos\left( \frac{n_i \pi
       y_i}{l_i}\right) \cos\left( \frac{n_i \pi x_i}{l_i} \right)
       \right) \left( \int_{0}^{t} I(\tau) \beta_{n_1,n_2}^{t - \tau}
       \ d\tau \right). \label{eqn:latest}
    \end{align}
    Define 
   \begin{align*}
     &\Theta'_{0}(t) = \frac{1}{l_1 l_2}\int_{0}^{t} I(\tau) \ d\tau,
     \nonumber \\ & P(y,n_1,n_2) = \prod_{i=1}^{2} \cos\left(
     \frac{n_i \pi y_i}{l_i}\right), \\ &A'(x,n_1,n_2) =
     \prod_{i=1}^{2} \cos\left( \frac{n_i \pi x_i}{l_i}\right), \\
     &\Theta'_{n_1,n_2}(t;x) = A'(x,n_1,n_2) \left( \int_{0}^{t}
     I(\tau) \beta_{n_1,n_2}^{t - \tau} \ d\tau \right).
   \end{align*}
   With this notation we can write (\ref{eqn:latest}) as
   \begin{align}
     & C(y,t;x) = \Theta'_{0}(t) + \sum_{n_1=1}^{\bar{n}_1}
     \sum_{n_2=1}^{\bar{n}_2} \Theta'_{n_1,n_2}(t;x) P(y,n_1,n_2).
     \label{eqn:rec}
   \end{align}
    From the function $I(t)$ in (\ref{eqn:intensity}), we have
   \begin{align*}
     \Theta'_{0}(k+1) =& \Theta'_{0}(k) + I(k+1), \\
     \Theta'_{\an{n_1,n_2}}(k+1;x) =& \beta_{n_1,n_2}
     \Theta'_{\an{n_1,n_2}}(k;x) + A'(x,n_1,n_2) I(k+1) \frac{\left(
     \beta_{n_1,n_2} - 1 \right)}{\log(\beta_{n_1,n_2})}.
   \end{align*}
   \an{Furthermore}, define
   \[
     \gamma_{n_1,n_2} = 1 + n_2  (n_1 - 1) + n_2.
   \]

   From (\ref{eqn:latest})~and~(\ref{eqn:rec}) we can write
   $C(y,k+1;x)$ as the output of the following state-space system:
   \begin{align}
     \Theta'(k+1;x) =& D' \Theta'(k;x) + B'(x) I(k+1) \nonumber \\
     C(y,k+1;x) =& H'(y) \Theta'(k+1;x), \label{eqn:ss2}
   \end{align}
   \an{where}
   \begin{align*}
       \Theta'_{k+1} =& \left[ \theta_0(k+1) \,\theta_{1,1}(k+1) \,
     \ldots\, \theta_{1,\bar{n}_2}(k+1) \, \theta_{2,1}(k+1) \,\ldots
     \, \theta_{\bar{n}_1,\bar{n}_2}(k+1)\right]^T,
   \end{align*}
   $D'$ is the diagonal matrix with
   \begin{align*}
     D'(1,1) = 1, \ \ \
     D'(\gamma_{n_1,n_2},\gamma_{n_1,n_2}) =
     \beta^m_{n_1,n_2},
   \end{align*}
   $B'(x)$ is the column vector
   \begin{align*}
     B'(x)\,(1,1) &= 1 \\ B'(x)\,(\gamma_{n_1,n_2},1) &= \frac{
     A'(x,n_1,n_2) \beta^m_{n_1,n_2} - 1}{\log(\beta_{n_1,n_2})},
   \end{align*}
   and $H'(y)$ is the row vector with
   \begin{align*}
     H'(y) &= \left[ 1 \, P(y,1,1) \, P(y,1,2) \, \ldots P(y,1,n_2) \,
       P(y,2,1) \, \ldots P(y,n_1,n_2) \right].
   \end{align*}
  From (\ref{eqn:gmp}), (\ref{eqn:ss2})~and~(\ref{eqn:measure}) \an{we have}
   \begin{align*}
     \left[
       \begin{array}{c}
	 \Theta'(k+1;x) \\
	 I(k+1)
       \end{array}
       \right]
     =&
     \left[
	 \begin{array}{cc}
	   D' & \rho B'(x) \\ 
	   0  & \rho
	 \end{array}
	 \right]
       \left[
       \begin{array}{c}
	 \Theta'(k;x) \\
	 I(k)
       \end{array}
       \right] + \left[
	 \begin{array}{c}
	   B'(x) \\
	   1
	   \end{array}
	 \right] S(k), \\
       R_{i}(k+1;x) = &  \left[
	 \begin{array}{cc}
	   H'(s_i) & 0
	 \end{array}
	 \right]\left[
       \begin{array}{c}
	 \Theta'(k+1;x) \\
	 I(k+1)
       \end{array}
       \right]
       + N_{i}(k+1).
   \end{align*}
   Thus, \an{the system dynamics are modeled} using a state-space
   model similar to (\ref{eqn:statespace}) that is parametrized by
   the unknown source location $x.$ Notice that in this case
   \[
   \Theta_i(k;x) = \Theta_j(k;x) =  \left[
       \begin{array}{c}
	 \Theta'(k;x) \\
	 I(k)
       \end{array}
       \right].
   \]
   \an{Hence,} complete information is available about the joint
   statistics of $\Theta_i(k;x)$ and $\Theta_j(k;x).$ If the sensors
   do not sense synchronously at the beginning of the slot, the model
   is more complex. Nevertheless, we can still identify a state vector
   and write the sensor measurements as the output process of a
   state-space system.

\end{document}